\documentclass[aps, pre, twocolumn, letter, superscriptaddress, longbibliography]{revtex4-1}

\newcommand{\mm}{\mathrm}

\usepackage{bm}

\usepackage{verbatim}
\usepackage[pdftex]{graphicx}
\usepackage{sidecap}
\usepackage[center]{subfigure}
\usepackage{rotating}
\usepackage[usenames]{xcolor}
\usepackage[normalem]{ulem}
\usepackage{amsmath}
\usepackage{hyperref}

\newcommand{\beginsupplement}{%
        \setcounter{table}{0}
        \renewcommand{\thetable}{S\arabic{table}}%
        \setcounter{figure}{0}
        \renewcommand{\thefigure}{S\arabic{figure}}%
     }

\begin{document}

\title{Jamming Criticality of Near-Crystals}

\author{Georgios Tsekenis}
\email{geotsek@gmail.com}
\affiliation{Dipartimento di Fisica, Sapienza Universita di Roma, Piazzale Aldo Moro 2, I-00185 Roma, Italy}
\affiliation{Department of Physics, University of Oregon, Eugene, Oregon 97403, USA}

\begin{abstract}
We report on the critical properties of minimaly-polydisperse crystals, hexagonal in 2d and face-centered cubic in 3 dimensions, at the isostatic jamming point. 
The force and gap distributions display power-law tails for small values. 
%The marginal stability relations of amorphous packings are violated.
The vibrational density of states (VDOS) is flat.
The scaling behavior of forces of extended floppy modes and the VDOS are universal and in agreement with an infinite-dimensional mean-field theory and maximally amorphous packings down to 2 dimensions. The distributions of gaps and forces of localized floppy modes of near-crystals appear non-universal. A small fraction of normal modes exhibit partial localization at low frequency. The majority of normal modes is delocalized exhibiting a characteristic inverse participation ratio scaling with frequency. The packing fraction and order at jamming decay linearly and quadratically respectively with polydispersity down to the maximally amorphous state.
%abstract current word count
\end{abstract}

\maketitle

{\it Introduction:} 
Jamming, prototypically, concerns the problem where a number of particles must fit in a certain volume, touch minimally without overlapping, and be globally rigid. At the jamming critical point there is no way to improve space coverage without violating at least one of the non-overlap constraints while one contact less makes the system loose. It represents a model for the rigidity transition of athermal matter~\cite{liu_nonlinear_1998, liu_jamming_2010, torquato_jammed_2010, parisi_mean-field_2010}.

Jamming criticality has been found in continuous constraint satisfaction problems (CCSP) such as the single layer perceptron~\cite{franz_universal_2015, franz_simplest_2016} and multilayer supervised learning models~\cite{franz_jamming_2019} (\cite{geiger_jamming_2019}). A transition point separates a SAT phase where all constraints are satisfied, from an unsatisfiable (UNSAT) phase. At the isostatic SAT-UNSAT point all variables are exactly constrained with the number of constraints equal to the number of variables.

At the packing jamming point the inter-particle contact network is isostatic as it can maintain mechanical equilibrium marginally. Randomness is necessary for isostaticity~\cite{maxwell_calculation_1864} to arise, since the crystal is also a packing solution of the jamming CCSP at a higher packing fraction~\cite{foot:xtal} than the maximally random jammed state~\cite{torquato_is_2000, torquato_jammed_2010, jiao_nonuniversality_2011, atkinson_existence_2014, torqua1}. 

Jamming has been recently connected with hard sphere glasses through a microscopic mean field theory (MFT) in the limit of infinite dimension~\cite{charbonneau_fractal_2014, charbonneau_glass_2017}. At infinite pressure a line of jamming points bounds the marginal glass phase, a.k.a. Gardner phase, where all levels of Replica Symmetry are broken (fullRSB) and the energy landscape is  fractal. 

Packings at the jamming line exhibit criticality with anomalous scaling exponents for the force and gap distribution~\cite{skoge_packing_2006,degiuli_force_2014, wyart_effects_2005, wyart_marginal_2012, lerner_low-energy_2013, charbonneau_universal_2012, charbonneau_fractal_2014, charbonneau_jamming_2015}, flat vibrational density of states (VDOS)~\cite{ohern_jamming_2003, franz_universal_2015} and characteristic normal modes that can exhibit quasi-localization~\cite{xu_anharmonic_2010, charbonneau_universal_2016}. Even though the problem is out of equilibrium where preparation protocol matters strongly~\cite{skoge_packing_2006, parisi_mean-field_2010} a jamming universality has emerged~\cite{charbonneau_fractal_2014, charbonneau_glass_2017} with an apparent very low upper critical dimension: critical exponents in computer experiments are the same down to $d=2$~\cite{charbonneau_jamming_2015}. 

\begin{figure}[t]
\centering
\includegraphics[scale=.99]{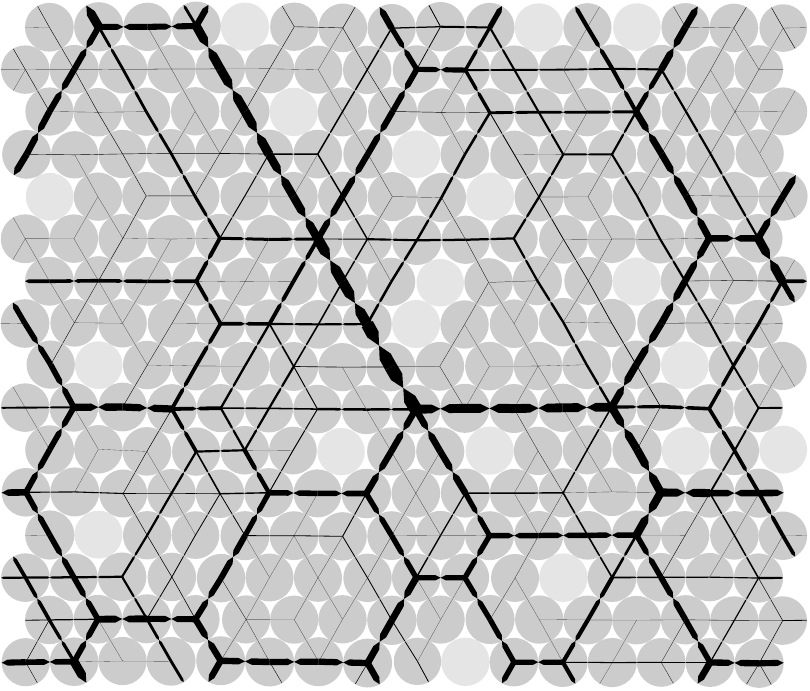}
\caption{ Visualization of a 2d HEX packing of $N=256$ particles at an isostatic energy minimum for polydispersity $s = 0.01$. $\varphi \approx 0.89801 \approx \varphi_{\mm{iso}} \approx 0.990 \varphi_{\mm{cp}}$. The force network of $N_{c} = 473 \equiv N_{\mm{c}, \mm{iso}}$ contacts is shown. The $N_{\mm{r}}=19$ \lq\lq lighter" particles are rattlers and experience zero force. All other particles bare at least $3$ $(=d+1)$ forces, some through the periodic boundary conditions. The magnitude of the forces is proportionally coded in the width of the vectors. Since the force distribution is power-law the smallest forces are shown equal so that they may be visible. (All symbols are defined in the text.)}
\label{fig:visualize2DHEX2}
\end{figure}

Monodisperse or polydisperse packings have been brought towards a jammed state with crystalline order either through energy minimization schemes of soft spheres~\cite{mari_jamming_2009, goodrich_solids_2014, tong_crystals_2015, charbonneau_glassy_2019} or as hard spheres with molecular dynamics~\cite{torquato_is_2000, torquato_jammed_2010} and linear programming optimization protocols~\cite{jiao_nonuniversality_2011, atkinson_existence_2014}.  

For the highly symmetric soft sphere crystals such as the hexagonal lattice (HEX) in 2d and the face-centered cubic (FCC) lattice in 3d with continuous polydispersity there is evidence of a phase of disordered crystals above the closed-packed density~\cite{tong_crystals_2015}. Some of the mechanical properties of disordered crystals approaching jamming, such as deformation moduli and contact number, exhibit numerical measurable scaling~\cite{tong_crystals_2015} that appears qualitatively consistent with an infinite-dimensional fullRSB model~\cite{ikeda2019jamming}. 

In this Letter we measure the critical jamming properties of slightly polydisperse crystals that correspond to the maximum-packing-fraction periodic structures in 2d and 3d, namely HEX and FCC~\cite{SI}. At their isostatic mechanical equilibrium these highly ordered near-crystals are described by characteristic scaling exponents and functions, a subset of which are in agreement with the corresponding quantities of amorphous packings in $d \geq 2$~\cite{charbonneau_jamming_2015} and the amorphous ($d=\infty$) MFT description~\cite{charbonneau_fractal_2014, charbonneau_glass_2017}. Another subset of exponents and exponent relations, however, do not appear to be classifiable under the known amoprhous jamming universality.

\begin{figure}[t]
\centering
\includegraphics[scale=.6]{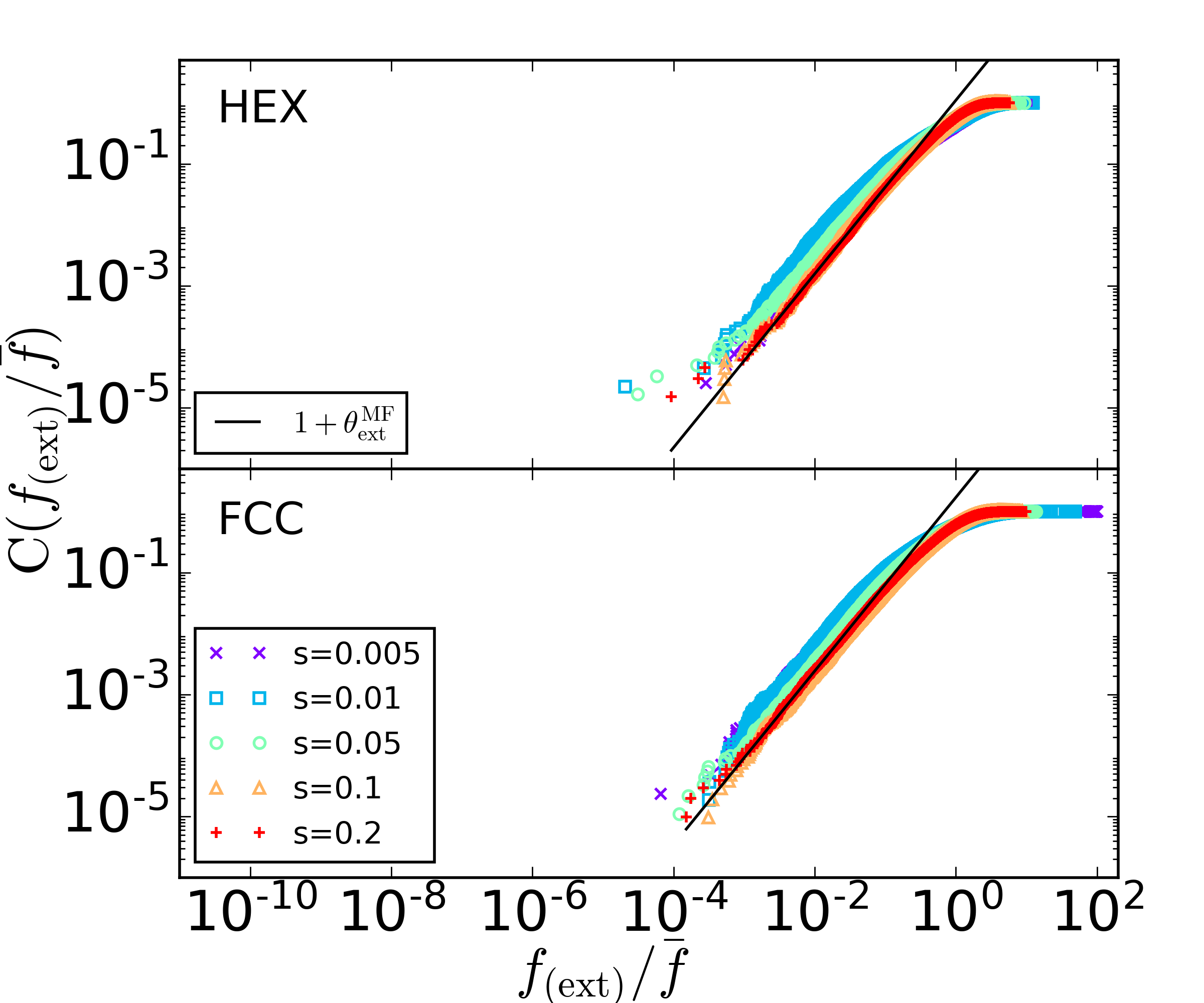}
\caption{Cumulative distribution of inter-particle forces that are related to an extended dipolar floppy mode for isostatic jammed near-crystals. For all polydispersities they exhibit power-law tails at small forces that scale as the mean-field solution of $1+\theta_{\mm{ext}}^{\mm{MF}}=1+0.42311$ (continuous line). 
%We define the cumulative distribution function as $G(x)=\int_{0}^{x}P(x')dx'$.
}
\label{fig:forcesExt}
\end{figure}

{\it Methodology:}
We arrange $N$ particles in a crystal structure in a commensurate cell~\cite{SI} with periodic boundary conditions (PBC). The particles interact via a soft harmonic potential only when they are in contact, $V=\sum_{\langle ij \rangle}\kappa(|\vec{r}_i-\vec{r}_j|-\sigma_{ij})^2\Theta(\sigma_{ij}-|\vec{r}_i-\vec{r}_j|)$, $\vec{r}_i$ is the coordinate vector of particle $i$, $\sigma_{ij}$ is the average diameter of particles $i$ and $j$, $\kappa$ is a stiffness and $\Theta$ the step function. We introduce a small amount of polydispersity $s$ by multiplying each diameter with a random number drawn from a lognormal distribution with unit mean and standard deviation $s$. 
Starting from a very high initial packing fraction~\cite{SI}, $\varphi_{0}=0.99$, we proceed to minimize the energy $V$ of the packing and shrink the radii of the particles in successive steps. The energy minimizations are performed with FIRE \cite{bitzek_structural_2006} on GPGPUs~\cite{pycud}. As the packing fraction is decreased a small number, $N_{\mm{r}}$, of particles called rattlers abandon the force network and are excluded from the analysis. The remaining $N'=N-N_{\mm{r}}$ particles are usually a few percent less than $N$. The minimum number of contacts (constraints) $N_{c}$ needed for the packing of $N'$ particles to be rigid is $N_{\mm{c}, \mm{iso}} \equiv (N'-1)d +1$ in PBC. At the initial packing fraction, $\varphi_{0}$, the system is over-constrained, $N_{c}>N_{\mm{c}, \mm{iso}}$. We follow the above procedure until the number of contacts satisfies the isostaticity condition, $N_{c}=N_{\mm{c}, \mm{iso}}$ at packing fraction $\varphi_{\mm{iso}}$. For harmonic spheres $V \propto (\varphi-\varphi_{\mm{J}})^2$~\cite{charbonneau_jamming_2015} which we fit to our sequence to get the packing fraction $\varphi_{\mm{J}}$ at the jamming point  (more details in ~\cite{charbonneau_jamming_2015} and Supplementary Material in there). In Fig.~\ref{fig:visualize2DHEX2} and Fig.~\ref{fig:vis2DHEX012345},~\ref{fig:vis3DFCC012} in the Supplementary Material (SM)~\cite{SI} we show examples of isostatic HEX and FCC near-crystals where for smaller polydispesities the near-crystals appear more crystalline. 
For polydispersity down to $s\approx10^{-4}$ our packings can reach isostaticity, enough computational effort afforded, even-though we may need to proceed with very small steps $\Delta \varphi$ of packing fraction as $\varphi \to \varphi_{\mm{iso}}$. For HEX(FCC) crystals in 2d(3d) we average over $N_{systems}=9$($18$) different runs of $N=4096$($2048$) particles in each run. 

\begin{figure}[t]
\centering
\includegraphics[scale=.6]{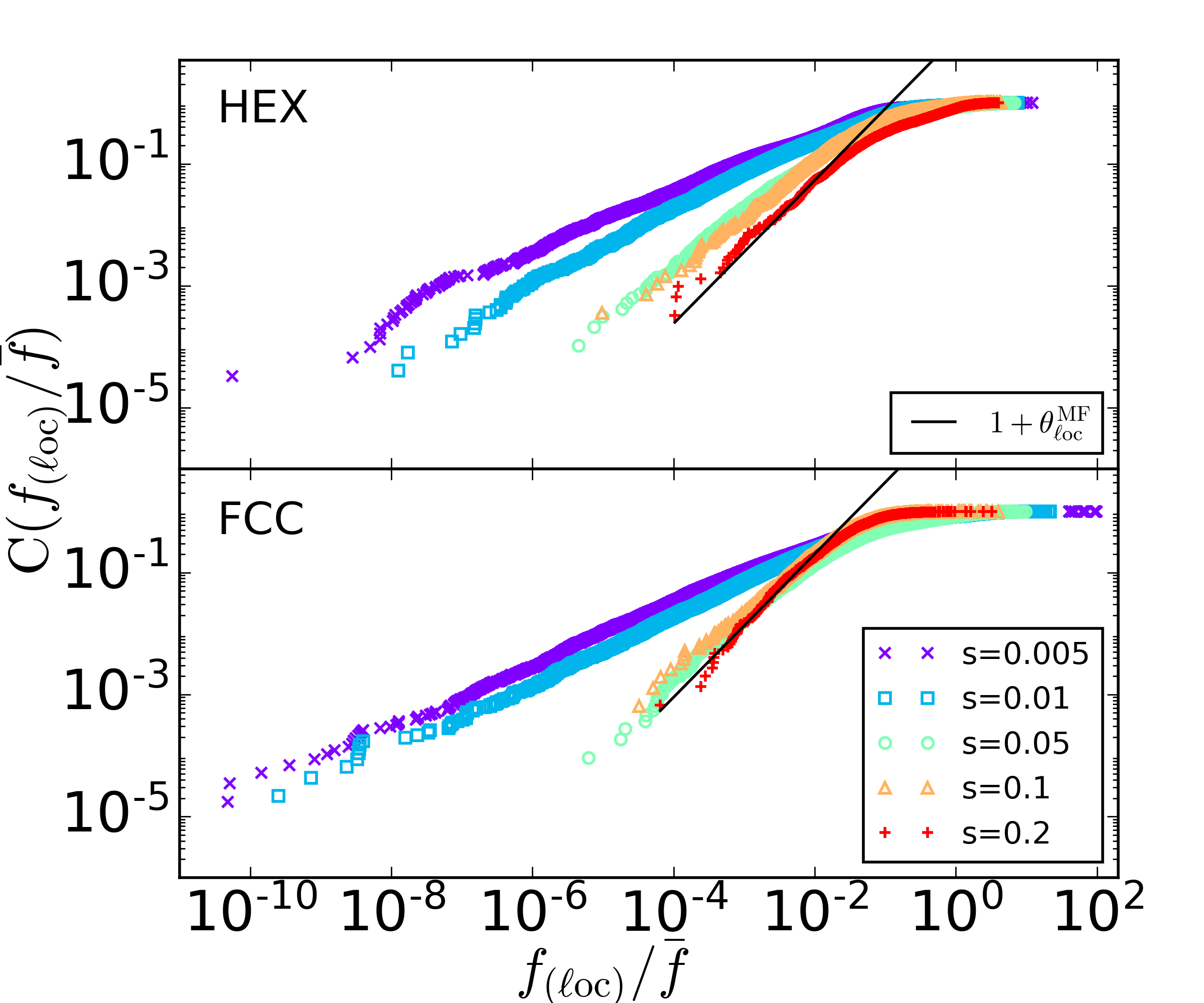}
\caption{Cumulative distribution of forces between particles that correspond to localized dipolar floppy modes for near-crystals at isostaticity. For small forces they have power-law tails with scaling exponents that increase with polydispersity reaching the MF value of $1+\theta_{\mm{\ell oc}}^{\mm{MF}}=1+0.17462$ (continuous line) at maximum amorphisation. 
%We define the cumulative distribution function as $G(x)=\int_{0}^{x}P(x')dx'$.
}
\label{fig:forcesLoc}
\end{figure}

\begin{figure}[t]
\centering
\includegraphics[scale=.6]{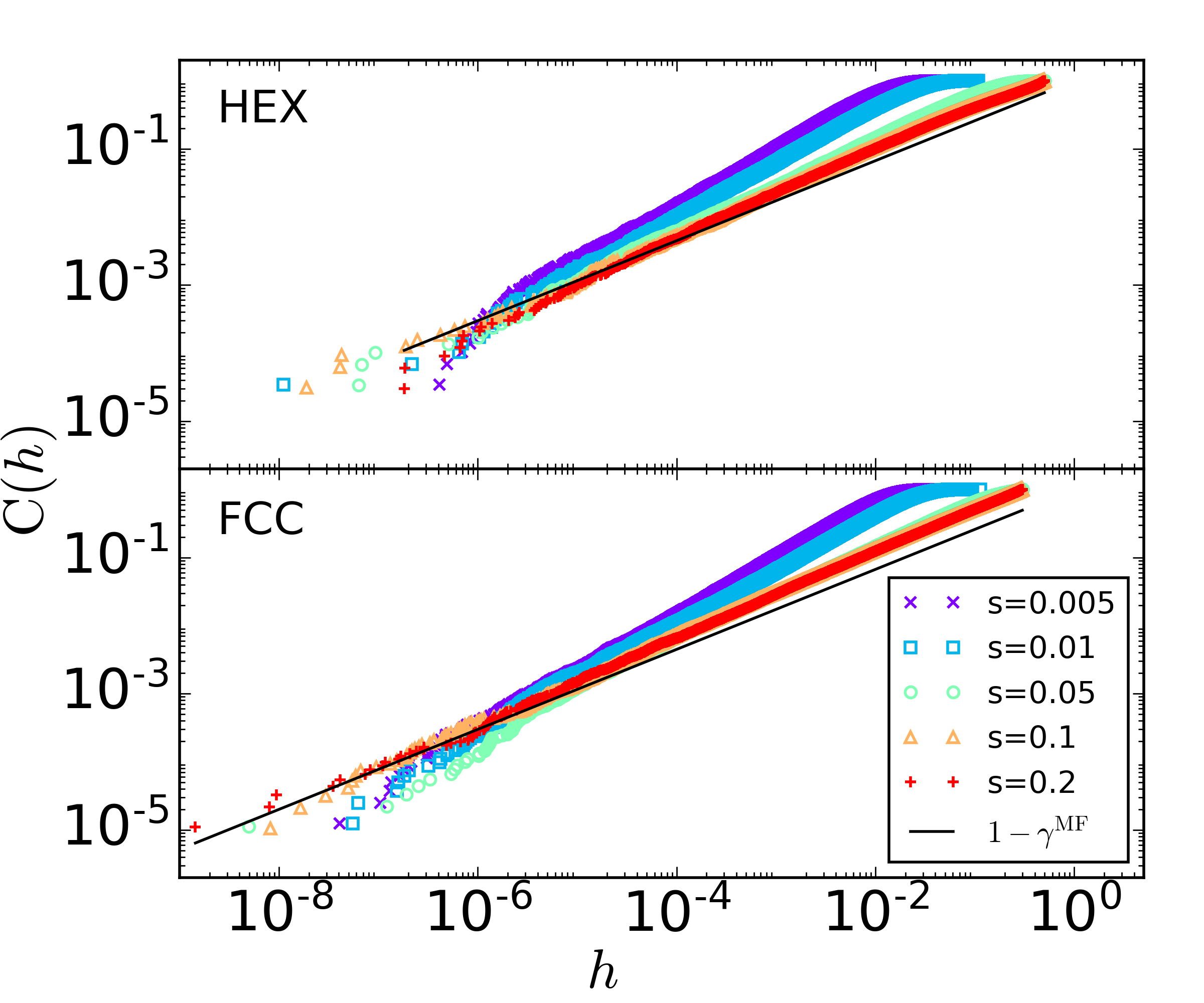}
\caption{Cumulative distribution of inter-particle gaps, i.e. the distances between neighboring non-touching particles for isostatic jammed near-crystals. The scaling exponent of the power-law small-value tails decreases with increasing polydispersity reaching the MFT value of $1-\gamma^{\mm{MF}}=1-0.41269$ (continuous line) at maximum amorphisation. The probability distribution of the same data is shown in Fig.~\ref{fig:gapsPDF}.
%We define the cumulative distribution function as $C(x)=\int_{0}^{x}D(x')dx'$.
}
\label{fig:gapsCDF}
\end{figure}

{\it Near-Crystal Jamming Properties:}
We calculate the inter-particle forces of the packing at the isostatic jamming point following the $S$-matrix construction (more details in ~\cite{charbonneau_jamming_2015} and Supplementary Material in there). The $S$-matrix is a matrix of contact unit vectors between the particles that are touching and it has dimensions of the $N_{c}$ contacts by the $N'd$ number of degrees of freedom. The symmetric matrix $\mathcal{N}=SS^{T}$ at isostaticity has a unique zero eigenvalue whose corresponding eigenvector contains the magnitudes of the inter-particle forces which comprise the random force network at mechanical equilibrium.

\begin{figure*}[]
\centering
\includegraphics[scale=.6]{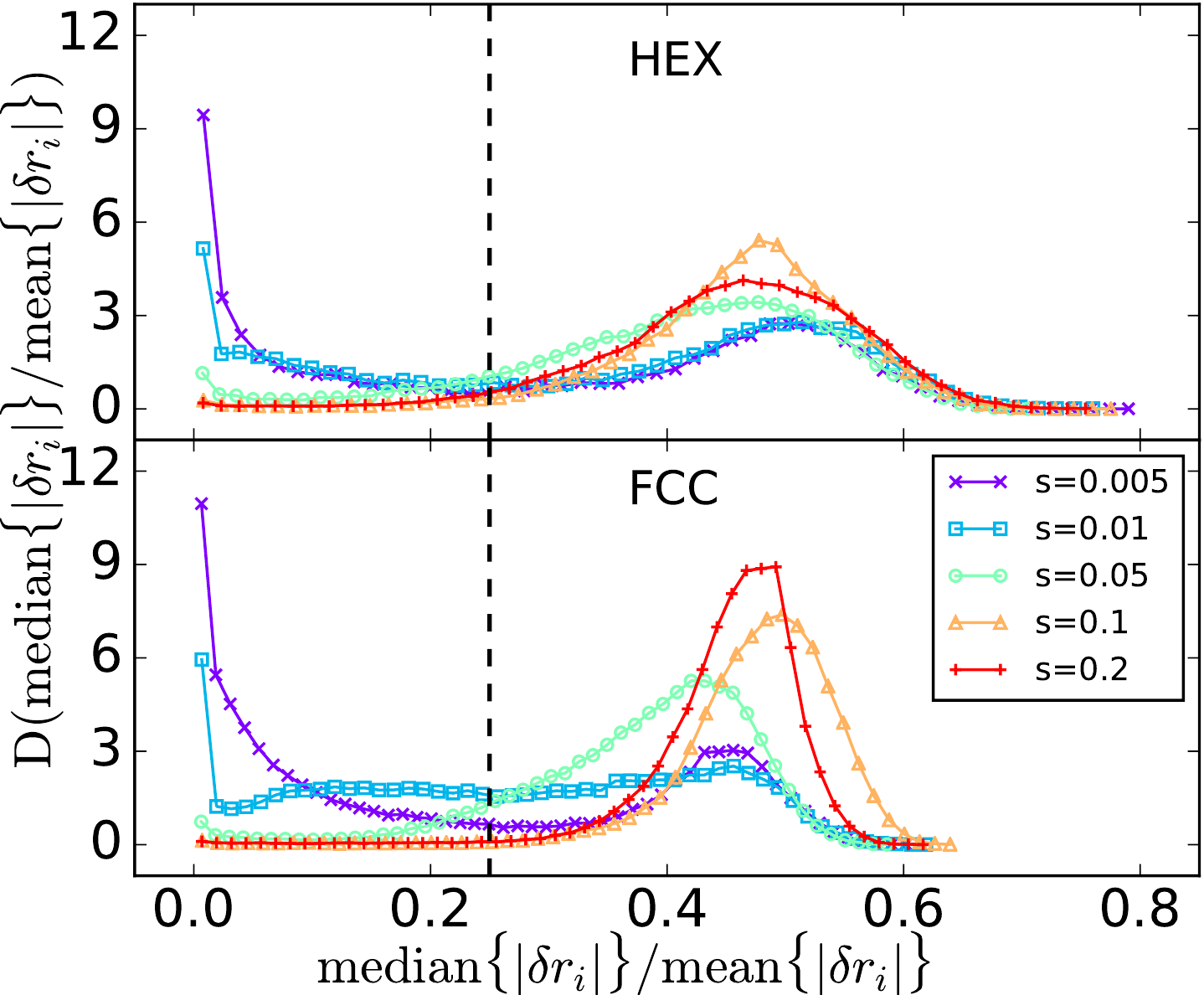}
\caption{The distribution of the median-over-mean of the particle displacements, $\delta \vec{r}$, of all dipolar floppy modes. The distinction between extended and localized dipolar floppy modes was taken at a threshold of $\mm{median} \{ |\delta r_i| \} / \mm{mean} \{ |\delta r_i| \}=0.25$ indicated by the vertical dashed lines roughly signifying the collective minimum of these distributions. Small variation in the threshold does not affect the results in Figs.~\ref{fig:forcesExt}, ~\ref{fig:forcesLoc} significantly. If contact $\langle ij \rangle$ is infinitesimally opened with a dipolar force then a corresponding floppy mode will appear in the jammed packing. The particle displacements will be given by the vector $\delta \vec{r}=H^{-1}S^{T}\vec{\tau}$ where $\vec{\tau}=\delta_{\tau,\langle ij \rangle}$ is a vector of a single $1$ and the rest of the entries $0$ indicating only contact $\langle ij \rangle$ was opened. $H = S^{T}S$ is the Hessian. All singular rows and columns of $S$ are eliminated in this procedure.}
\label{fig:floppyModesPDF}
\end{figure*}

\begin{figure*}[]
\centering
\includegraphics[scale=.55]{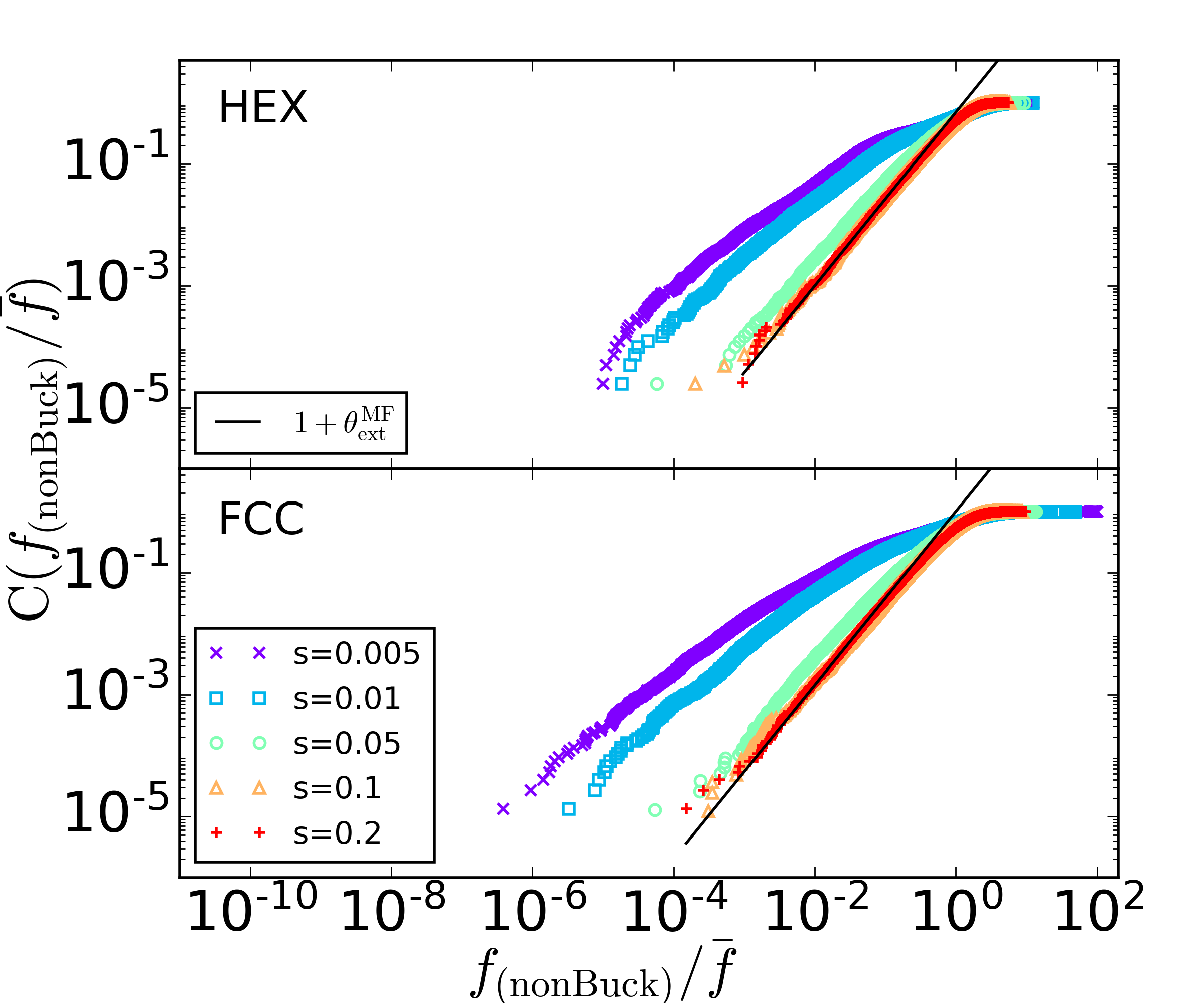}
\includegraphics[scale=.55]{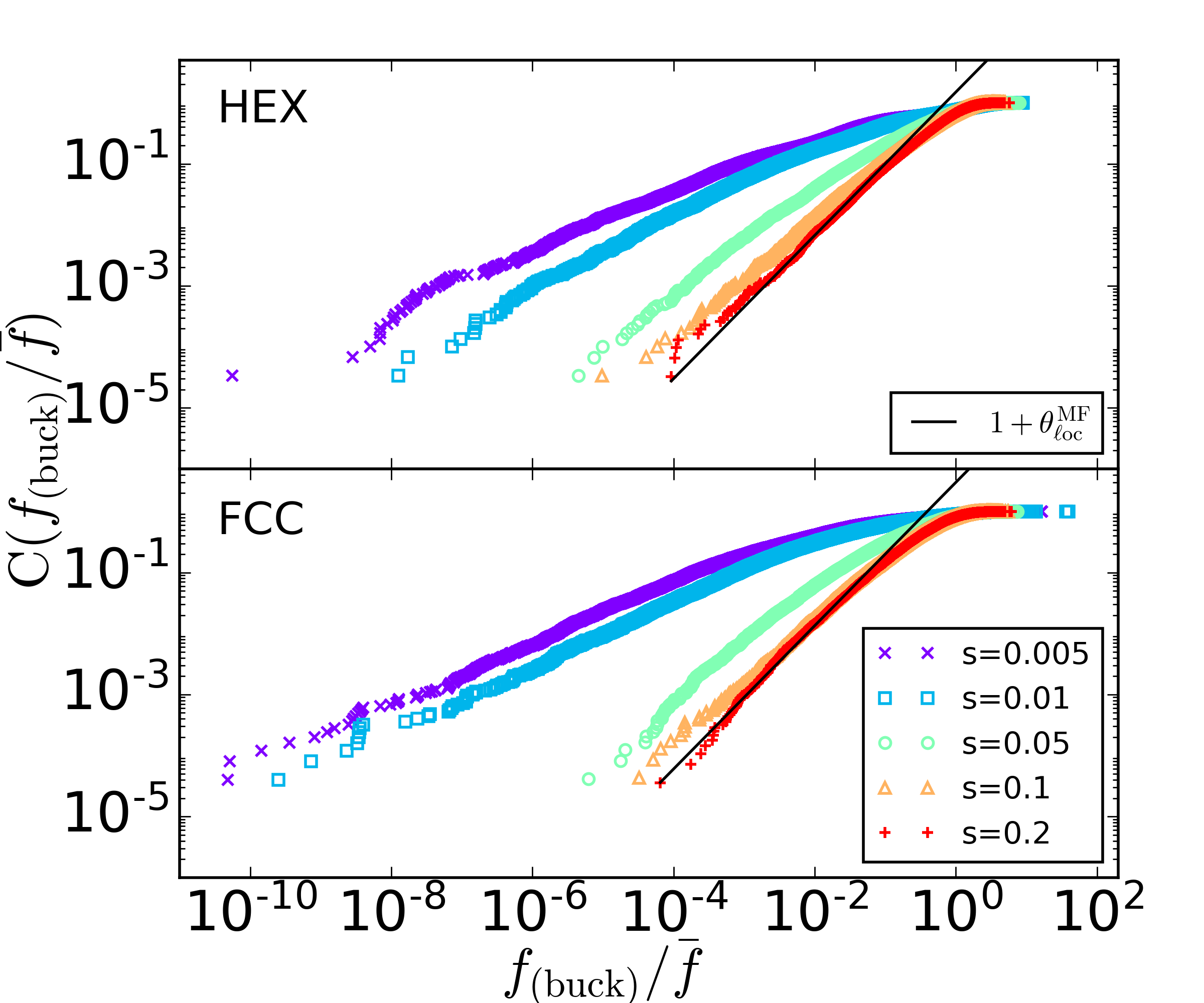}
\caption{ (left) Cumulative distribution of forces on non-buckler particles. For larger polydispersities they exhibit power-law tails that align well with the mean-field exponent of $1+\theta^{\mm{MF}}_{\mm{ext}}=1+0.42311$. For smaller polydispersities the distributions have not saturated to a power law in contrast with the extended forces based on the dipolar floppy modes separation, Fig.~\ref{fig:forcesExt}. \\
(right) Cumulative distribution of forces on buckler particles. For larger polydispersities they exhibit power-law tails that align well with the mean-field exponent of $1+\theta^{\mm{MF}}_{\mm{\ell oc}}=1+0.17462$. For smaller polydispersities they appear to follow power laws with smaller exponents reminiscent of the scaling exhibited by the localized forces based on the dipolar floppy modes characterization, Fig.~\ref{fig:forcesLoc}. A buckler particle is a stable particle baring $d+1$ forces.}
\label{fig:forcesBuckNonBuck}
\end{figure*}

The distributions of the force magnitudes exhibit power-law tails at small values (Eqs.~\ref{eqs:cdfext}, ~\ref{eqs:cdfloc}) a feature also seen in maximally amorphous packings ~\cite{lerner_low-energy_2013, degiuli_force_2014, charbonneau_fractal_2014, charbonneau_jamming_2015, kallus_scaling_2016, franz_simplest_2016}. In Fig.~\ref{fig:visualize2DHEX2} and Fig.~\ref{fig:vis2DHEX012345} in SM~\cite{SI} one can see that the force network of isostatic near-crystals is disordered even though the underlying structure may be highly ordered. We use the dipolar floppy modes~\cite{muller_marginal_2015} to separate the forces between those whose contact is related to an extended as opposed to a localized dipolar excitation (more details in~\cite{charbonneau_jamming_2015, charbonneau_glassy_2019} and Supplementary Material in there). An infinitesimal opening of a contact with a force dipole will induce an elementary excitation in the packing, which, if all other contacts stay the same, will not change the energy of the system. Such a dipolar floppy mode is termed extended (localized) if a large (small) fraction of particles moves appreciably~\cite{muller_marginal_2015}. In Fig.~\ref{fig:floppyModesPDF} we plot the distribution of the median over the mean of particle displacements, $\delta \vec{r}$, of the dipolar floppy mode of each contact~\cite{SI}. Localized (Extended) forces correspond to dipolar floppy modes with $\mm{median} \{ |\delta r_i| \} / \mm{mean} \{ |\delta r_i| \}$ below (above) a threshold value (see Fig.~\ref{fig:floppyModesPDF}) and their cumulative distribution is shown in Fig.~\ref{fig:forcesExt} and Fig.~\ref{fig:forcesLoc}. (The non-buckler vs buckler characterization of forces as in maximally amorphous packings~\cite{charbonneau_jamming_2015} gives similar results, shown in Fig.~\ref{fig:forcesBuckNonBuck} in SM~\cite{SI}.) The distributions of extended forces, Fig.~\ref{fig:forcesExt}, in minimally disordered crystals for all $s$ and both $d=2,3$ are in agreement with the $d=\infty$ MFT result~\cite{charbonneau_fractal_2014, charbonneau_glass_2017}. However an exponent that decreases with decreasing polydispersity is revealed for the localized-force distributions, Fig.~\ref{fig:forcesLoc}, which tends to the MFT values for high enough amorphisation.
% as the amorphisation transition is approached.

\begin{figure*}[]
\centering
\includegraphics[scale=.6]{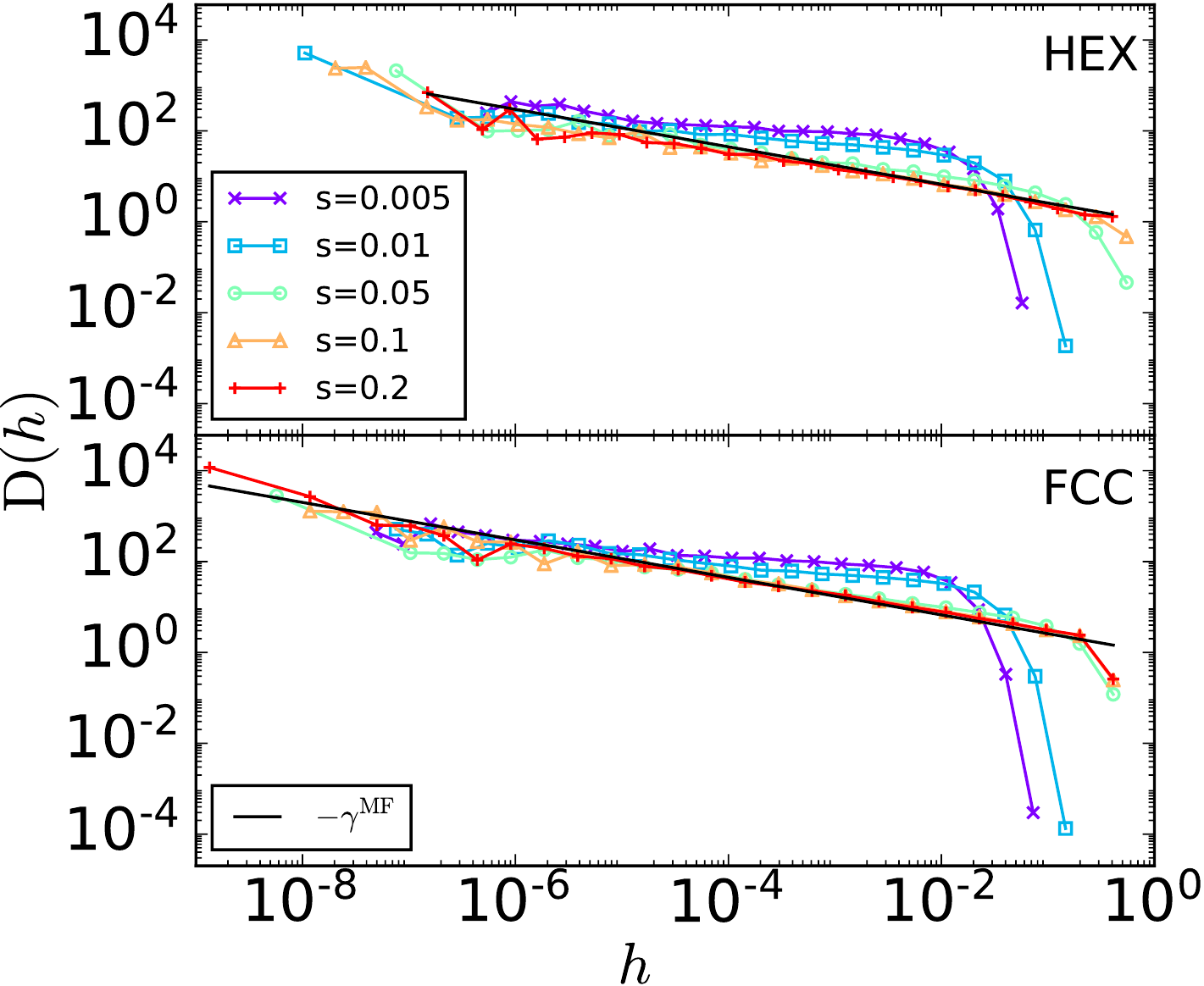}
\caption{ Distribution of inter-particle gaps, i.e. the distances between neighboring non-touching particles.  The MFT value of $\gamma^{\mm{MF}}=0.41269$ (continuous line) is approached by the distributions as polydispersity is increased. Same data as in the cumulative distribution plot of Fig.~\ref{fig:gapsCDF}.}
\label{fig:gapsPDF}
\end{figure*}

The distribution of the gaps, $h = r_{ij}/\sigma_{ij}-1$, between neighboring but non-touching particles, is power-law for minimally disordered crystals at jamming as in maximally amorphous systems (Eq.~\ref{eqs:cdfgap}). The scaling exponent, $\gamma$, for small gaps increases with polydispersity in near-crystals (Fig.~\ref{fig:gapsCDF} and~\ref{fig:gapsPDF}) tending to the MFT value~\cite{skoge_packing_2006, charbonneau_universal_2012, lerner_low-energy_2013, charbonneau_fractal_2014, charbonneau_jamming_2015, charbonneau_glass_2017} of amorphous jamming at high polydispersity.
 
\begin{eqnarray}
\label{eqs:cdfext}
\mm{C}(f_{(\mm{ext})}) & \sim & f_{(\mm{ext})}^{1+\theta_{\mm{ext}}}\\
\label{eqs:cdfloc}
\mm{C}(f_{(\mm{\ell oc})}) & \sim & f_{(\mm{\ell oc})}^{1+\theta_{\mm{\ell oc}}}\\
\mm{C}(h) & \sim & h^{1-\gamma}.
\label{eqs:cdfgap}
\end{eqnarray}

The exponent relations that signify mechanical stability for amorphous solids, $1/(2+\theta_{\mm{ext}}) \leq \gamma$ and $1-\theta_{\mm{\ell oc}} \leq 2\gamma$ ~\cite{wyart_marginal_2012, lerner_low-energy_2013, charbonneau_fractal_2014, muller_marginal_2015, franz_simplest_2016}, are violated for isostatic disordered crystals. It seems that the long-range structural correlations may need to be taken into account in a new set of inequalities for the marginal mechanical stability of jammed near-crystals~\cite{charbonneau_glassy_2019}.

\begin{figure}[t]
\centering
\includegraphics[scale=.6]{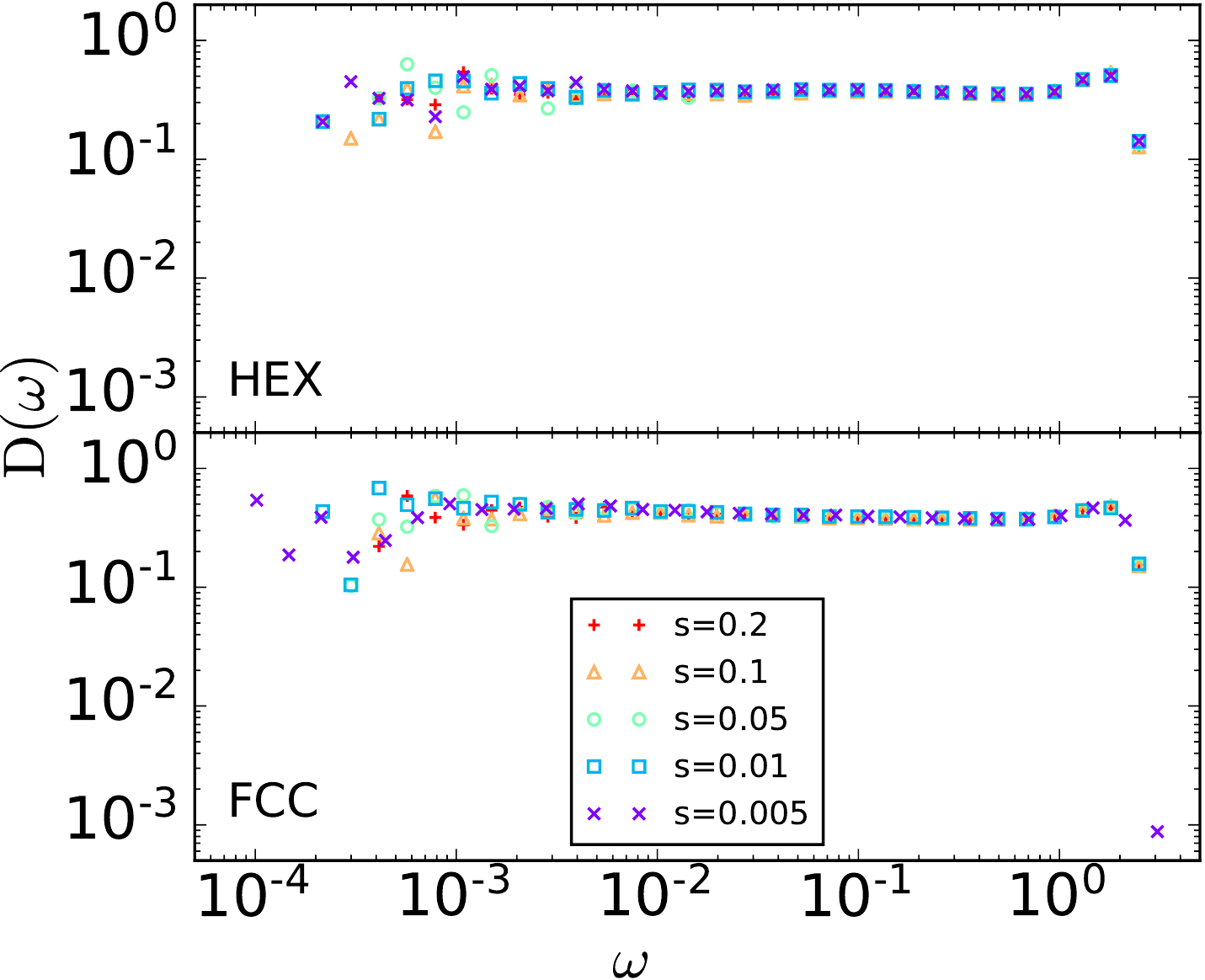}
\caption{ The distribution of vibrational frequencies (VDOS) appear constant as $\omega \to 0$ for isostatic near-crystals and clearly distinct from Debye $D(\omega) \sim \omega^{d-1}$ for all amounts of order.
}
\label{fig:VDOS}
\end{figure}

The Hessian is the matrix of curvature in configurational space at the isostatic energy minima (inherent structures) that our packings arrive. In general it can be calculated in Euclidean space as the matrix of second spatial derivatives of the potential energy of the packing  
\begin{eqnarray}
H_{ij} = \frac{\partial V^2}{\partial{x_{i} }\partial{x_{j} }} = (S^{T}S)_{ij}.
\end{eqnarray}
At the jamming point the Hessian can be expressed through the $S$-matrix of contact vectors which is commonly called a geometric construction and is identical to the above expression at isostaticity for the harmonic interaction. The eigenvalues $\lambda$ and eigenvectors $\vec{u}(\lambda)$ of $H$ give us the frequency $\omega = \sqrt{\lambda}$ and normal modes of the packing~\cite{ohern_jamming_2003, silbert_vibrations_2005, franz_universal_2015, charbonneau_universal_2016}. The $d$ zero-frequency modes correspond to the trivial global translations of the packing and are ignored in this analysis. The density of vibrational states for the near-crystal packing appear flat (Fig.~\ref{fig:VDOS}) in both $d=2,3$ and clearly distinct from Debye $D(\omega) \sim \omega^{d-1}$ ~\cite{mari_jamming_2009, tong_crystals_2015, charbonneau_glassy_2019}.

\begin{figure}[t]
\centering
\includegraphics[scale=.6]{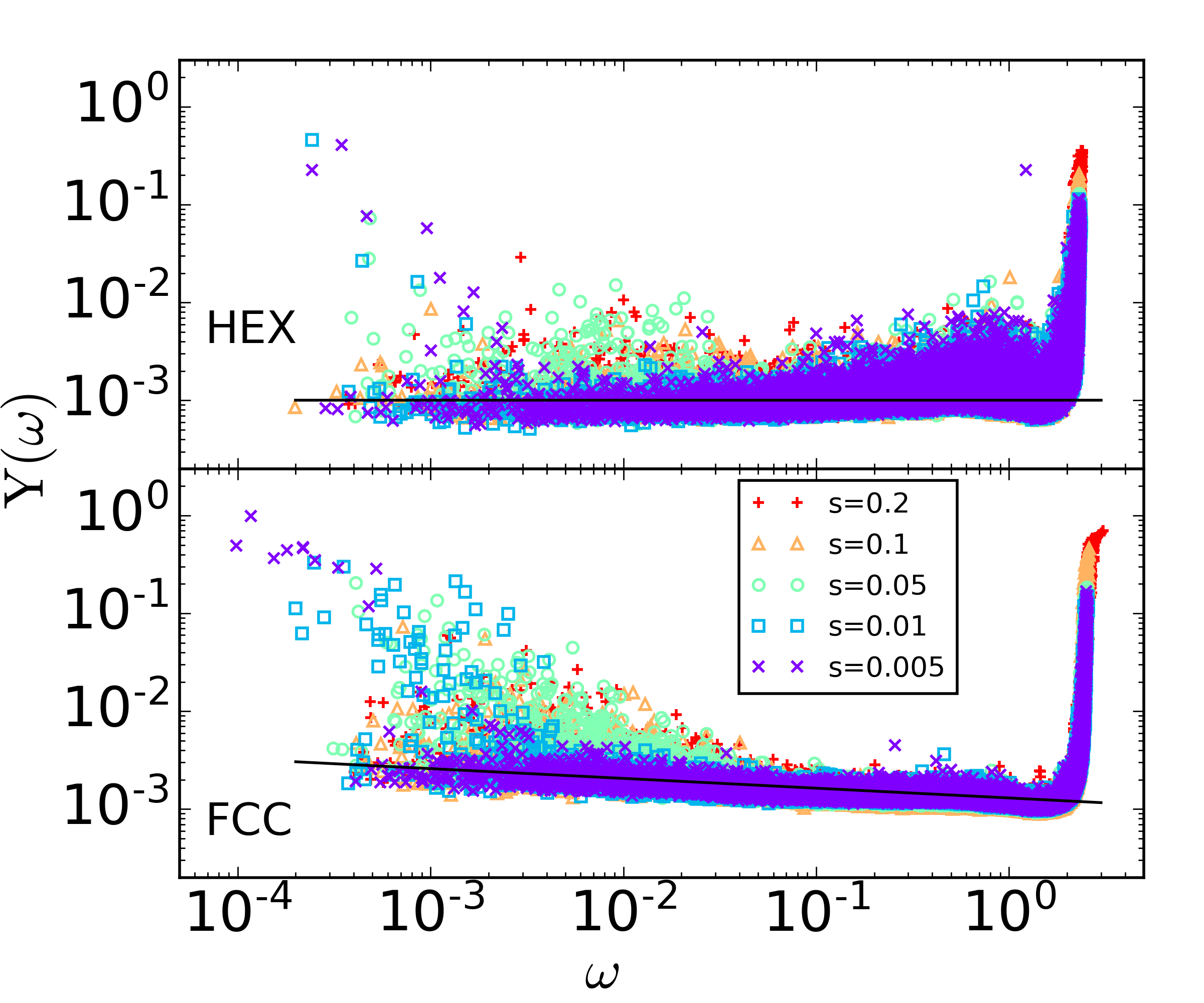}
\caption{ The inverse participation ratio as a function of frequency for near-crystals at the isostatic jamming point. For smaller frequencies and polydispersities the crystal packings exhibit increasing localization for a small fraction of their normal modes. The main delocalized branch scales as $Y(\omega) \sim \omega^{-0.1}$ for FCC, and is flat for HEX (continuous lines). In Figs.~\ref{fig:nmodes2DHEX-A},\ref{fig:nmodes3DFCC-A} we show the smallest and largest IPR normal mode for each polyispersity and crystal.
}
\label{fig:IPR}
\end{figure}

\begin{figure*}[]
\centering
\includegraphics[scale=.59]{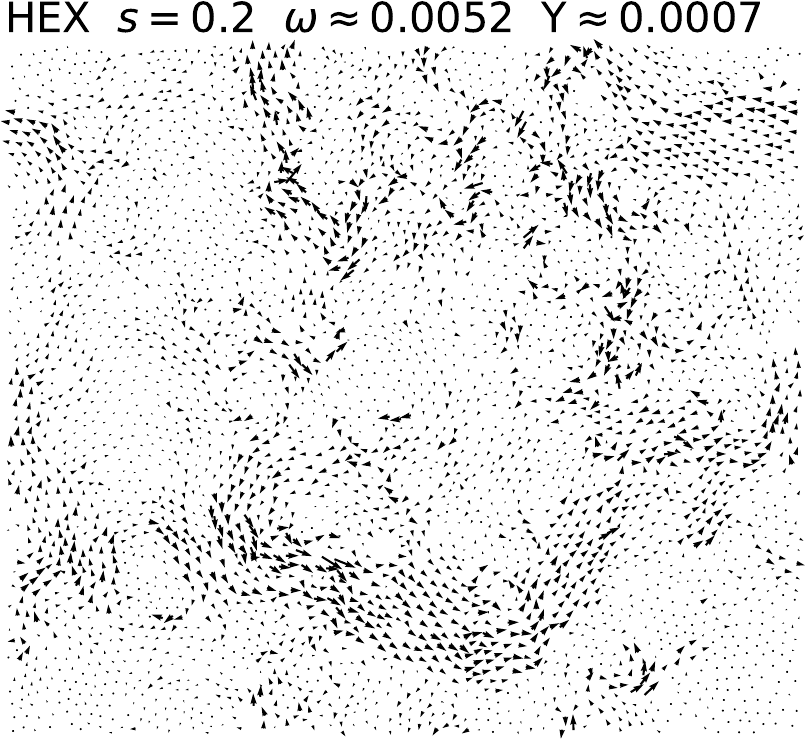}
\includegraphics[scale=.59]{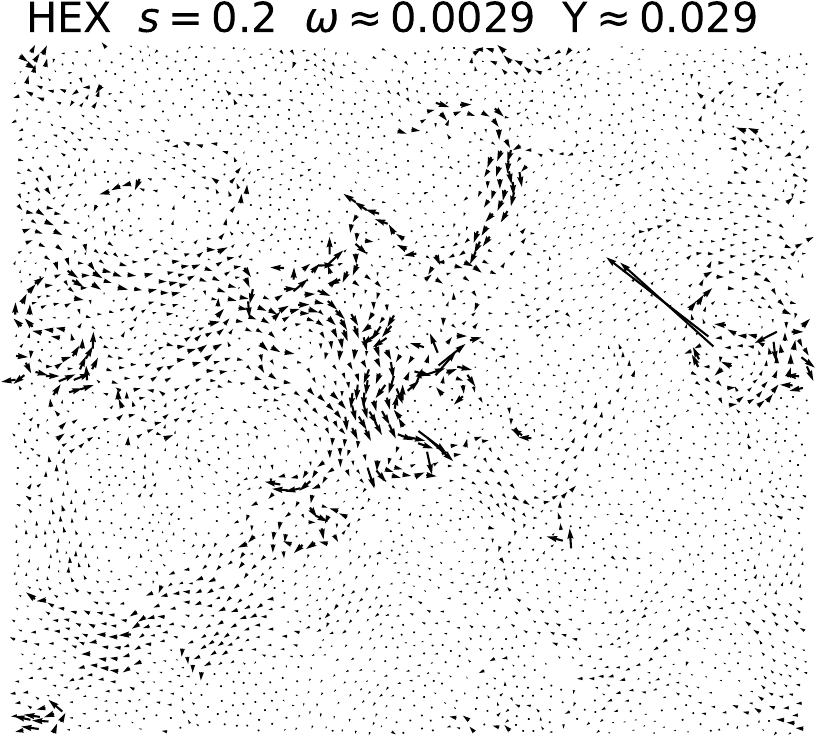}\\
\includegraphics[scale=.59]{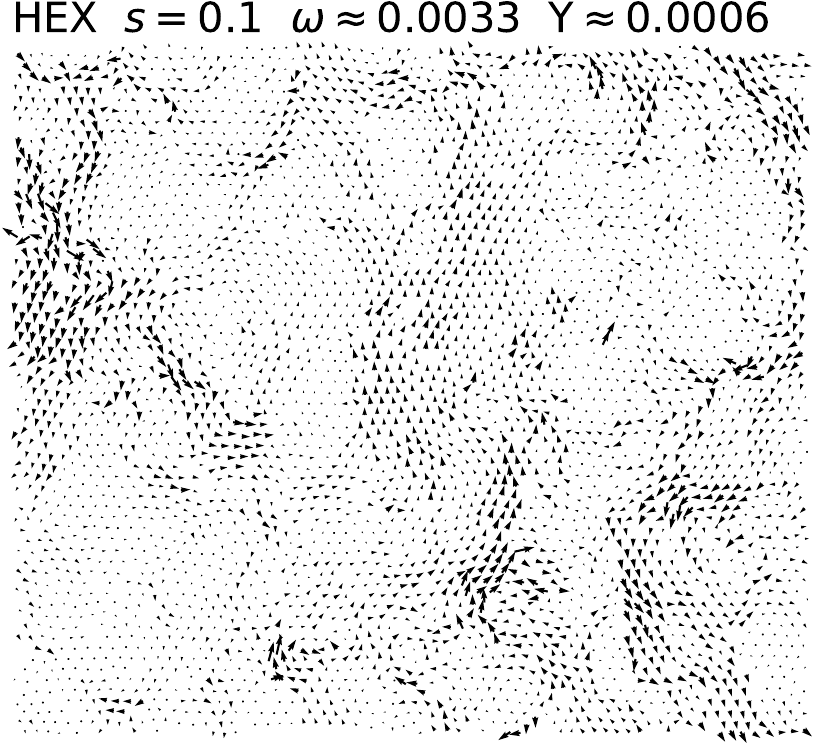}
\includegraphics[scale=.59]{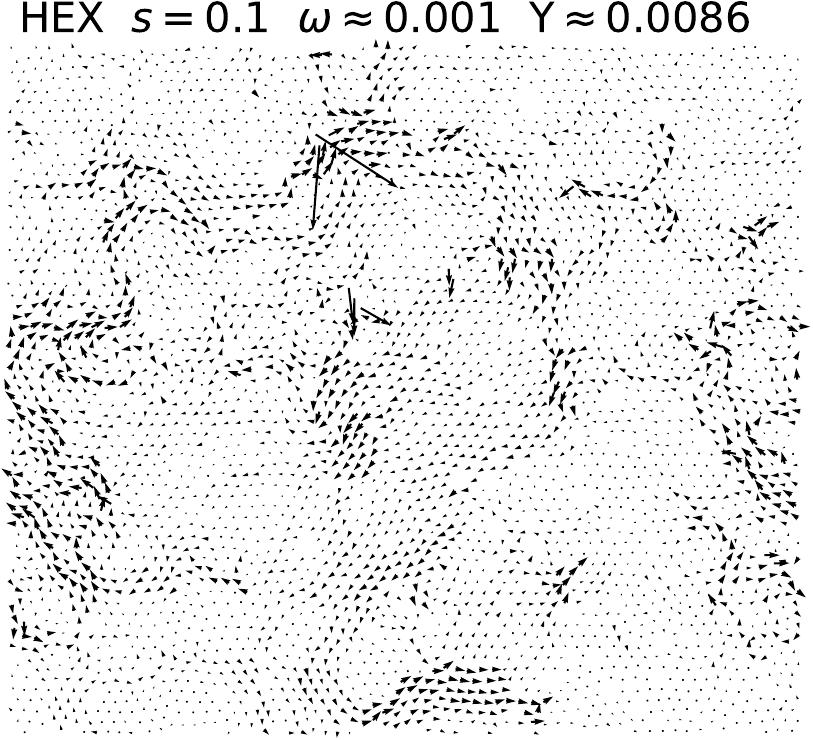}\\
\includegraphics[scale=.59]{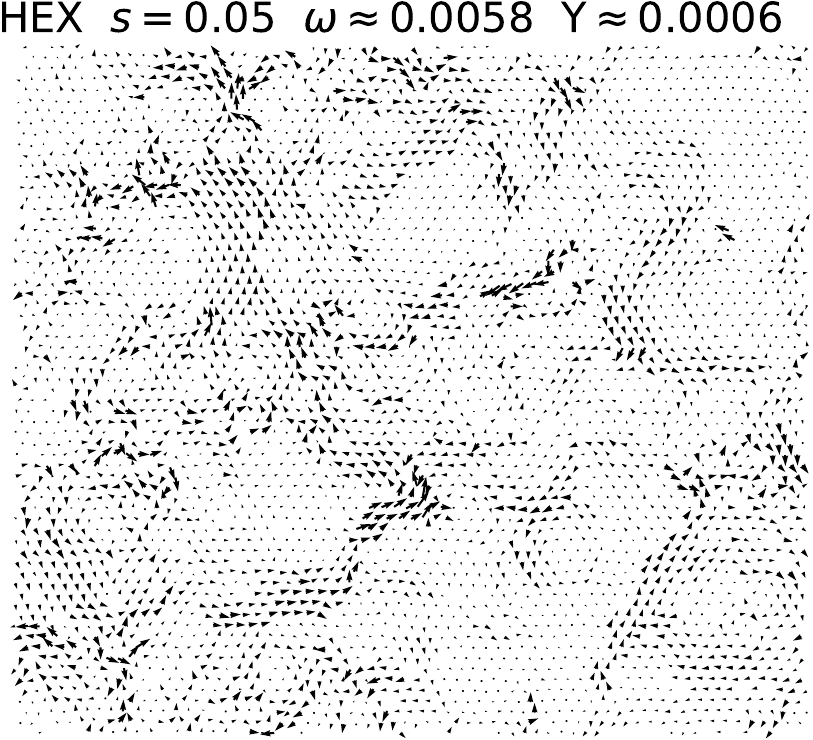}
\includegraphics[scale=.59]{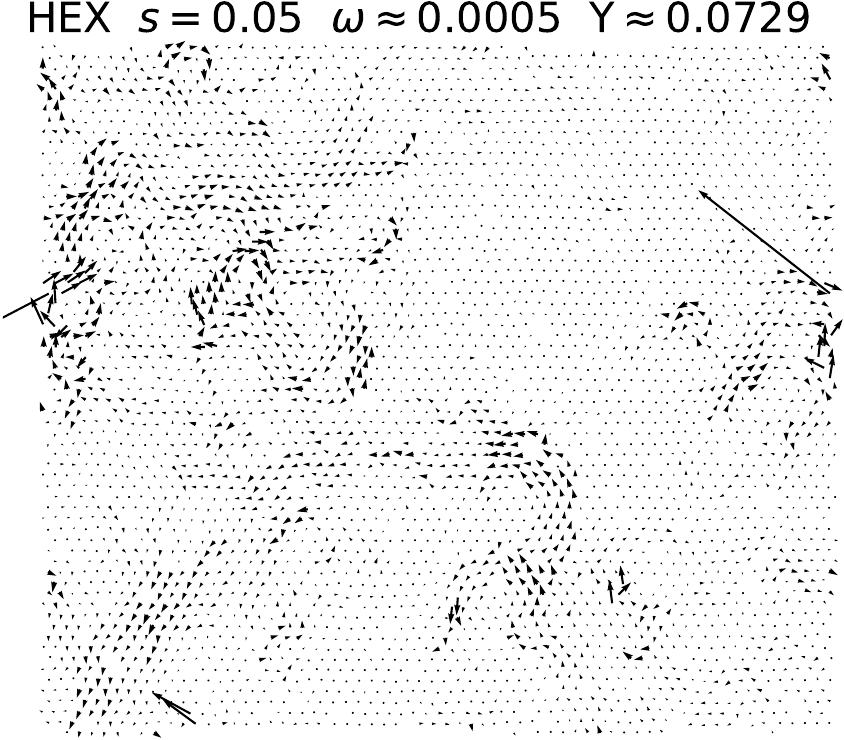}\\
\includegraphics[scale=.59]{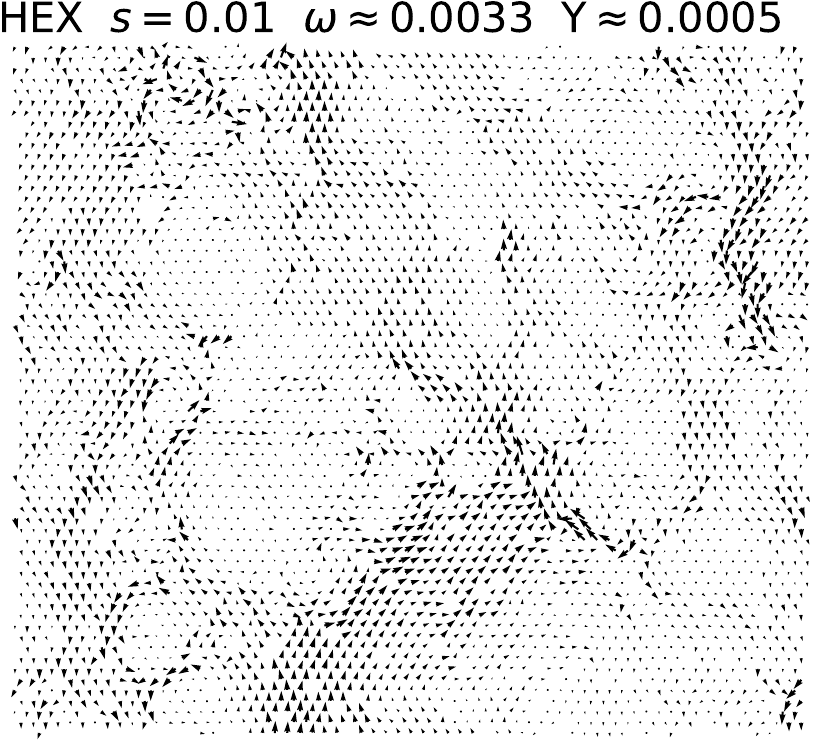}
\includegraphics[scale=.59]{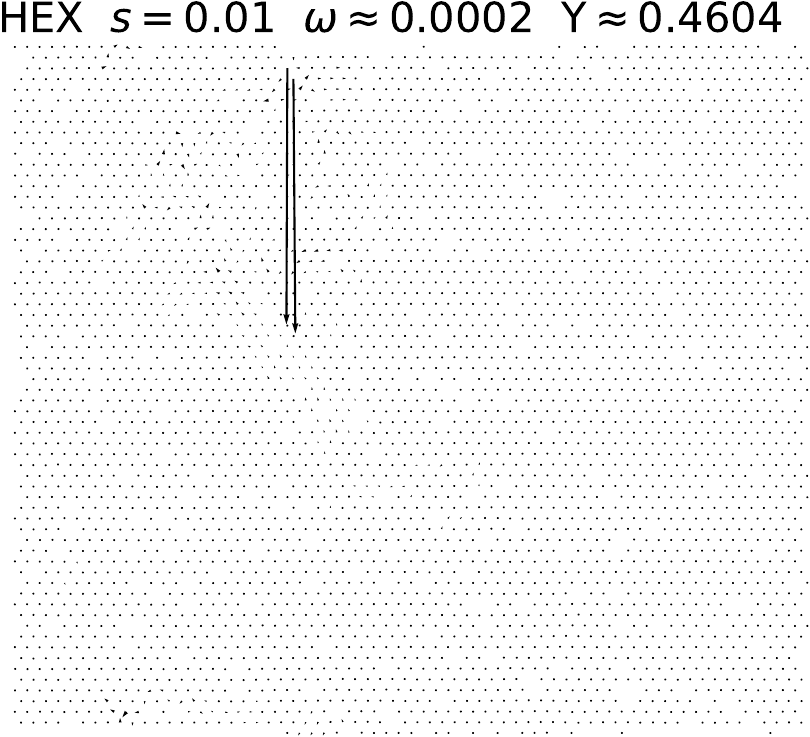}\\
\includegraphics[scale=.59]{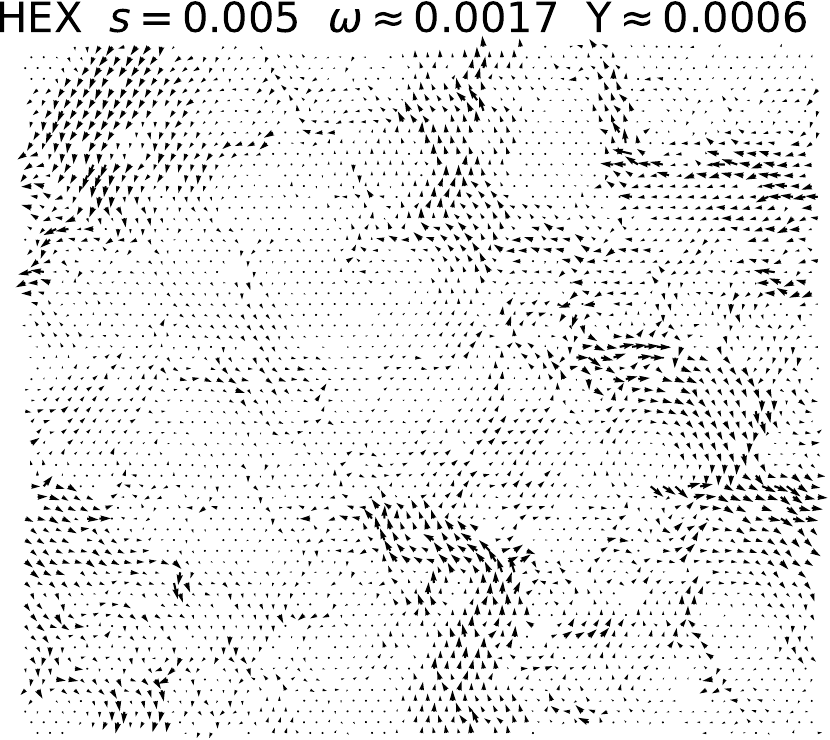}
\includegraphics[scale=.59]{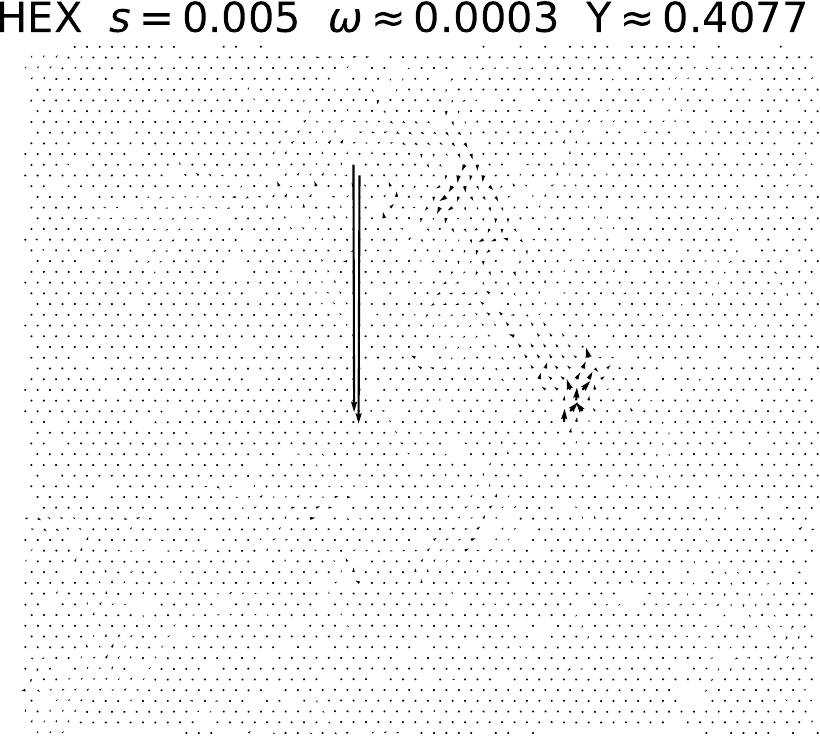}
\caption{ Plot of one of the smallest- (left column) and largest- (right column) IPR normal modes of a jammed 2d HEX near-crystal of $N = 4096$ particles for each polydispersity $s$. The frequency $\omega$ and the IPR $\mm{Y}$ are given (Fig.~\ref{fig:IPR}).
}
\label{fig:nmodes2DHEX-A}
\end{figure*}

\begin{figure*}[]
\centering
\includegraphics[scale=.65]{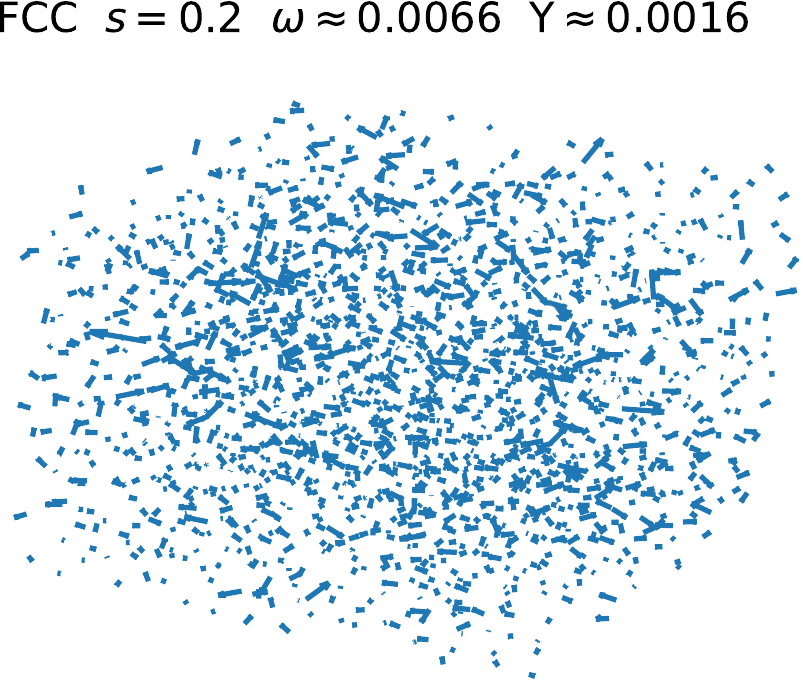}
\includegraphics[scale=.65]{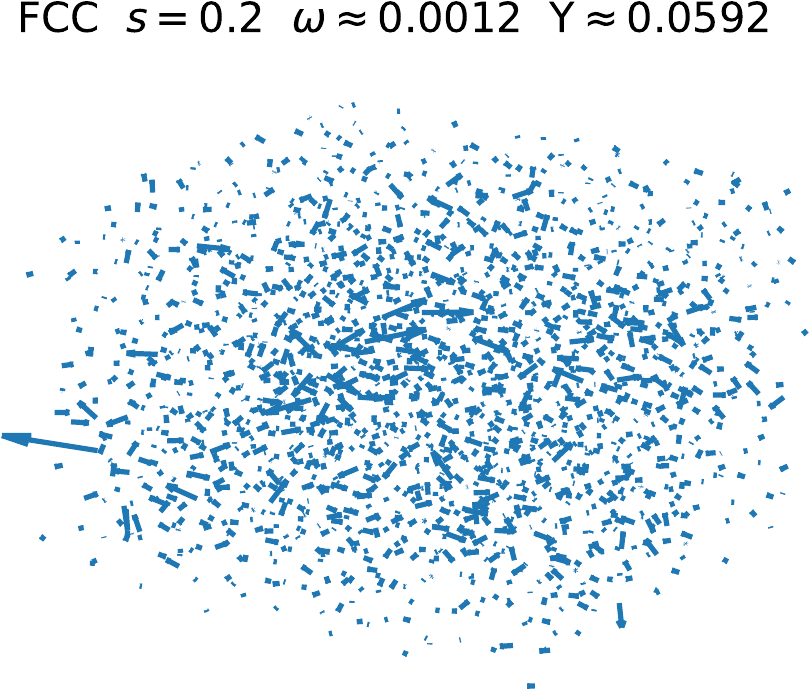}\\
\includegraphics[scale=.65]{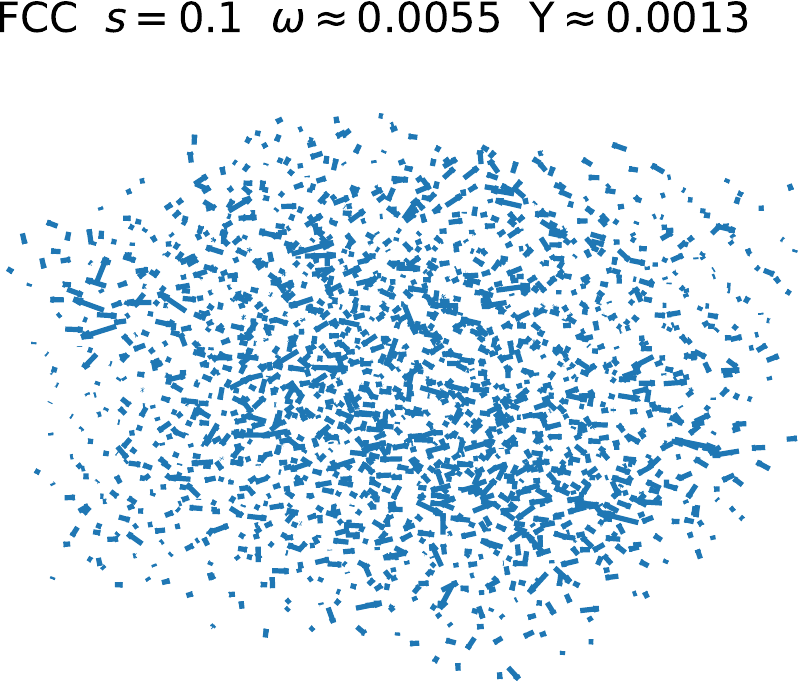}
\includegraphics[scale=.65]{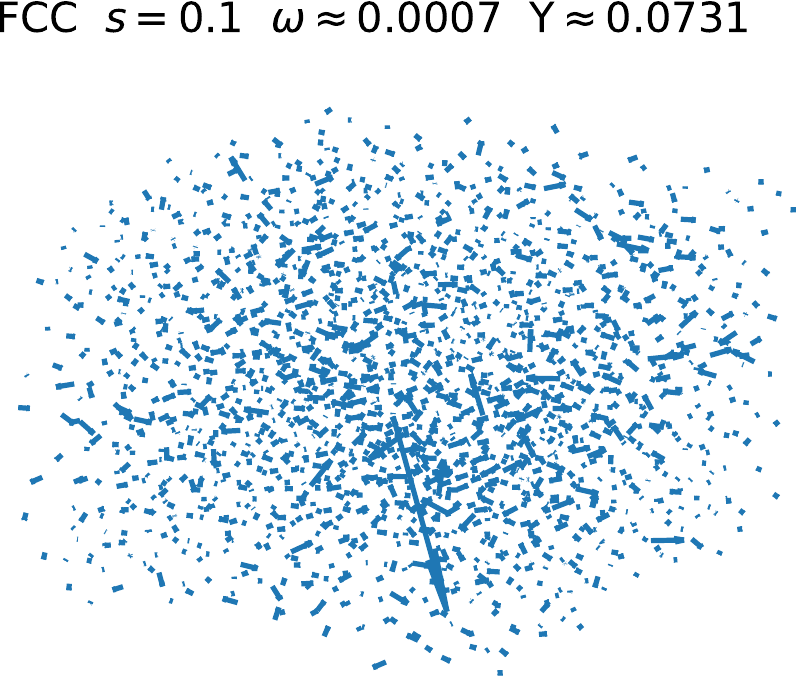}\\
\includegraphics[scale=.65]{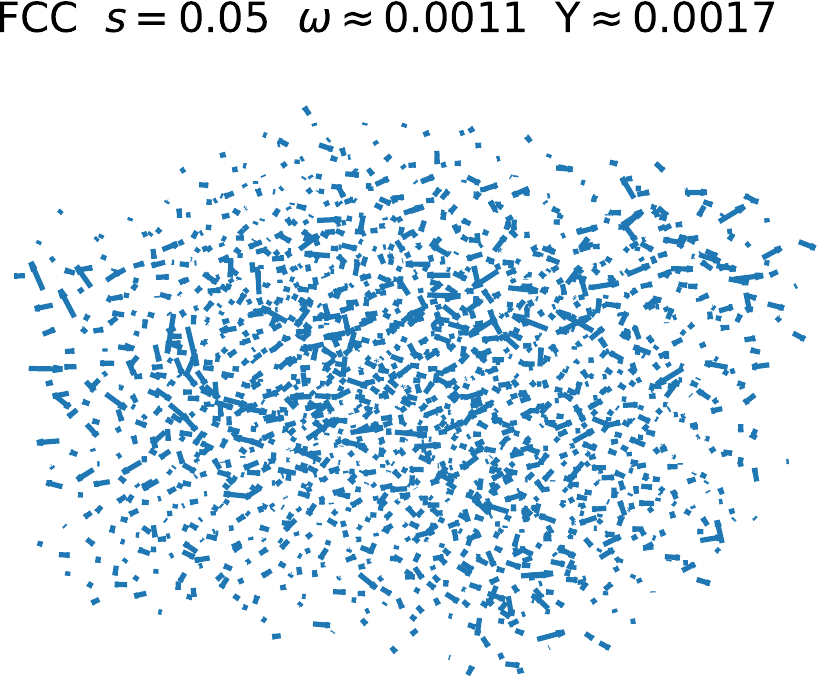}
\includegraphics[scale=.65]{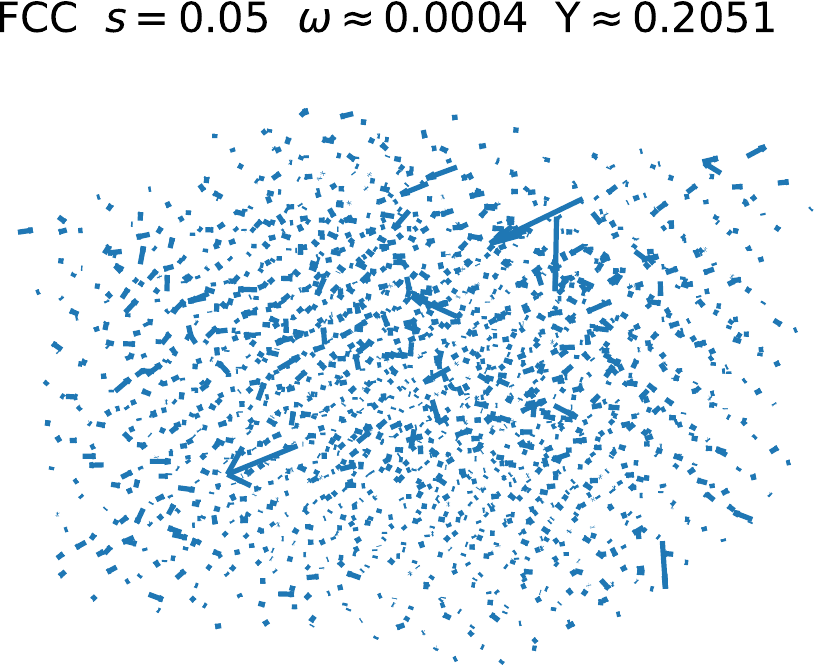}\\
\includegraphics[scale=.65]{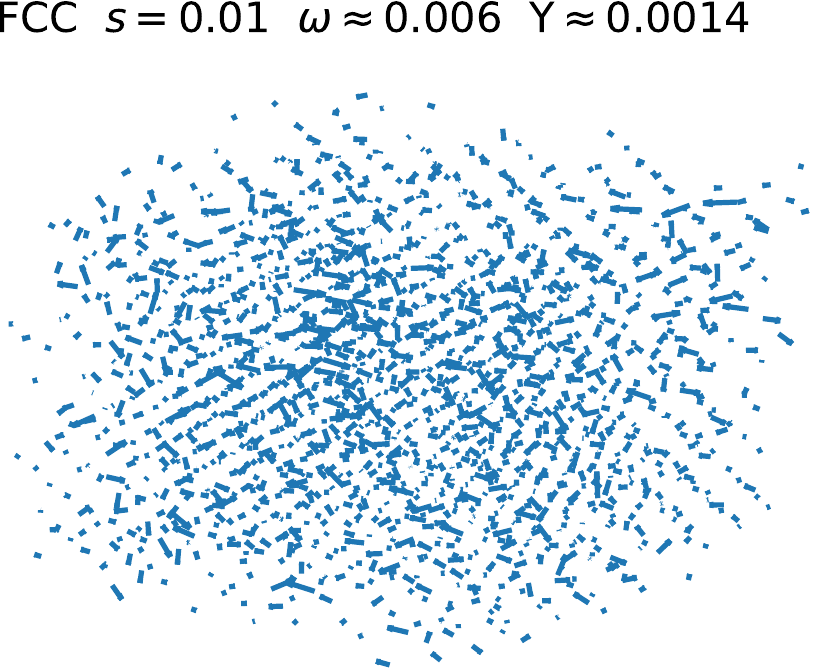}
\includegraphics[scale=.65]{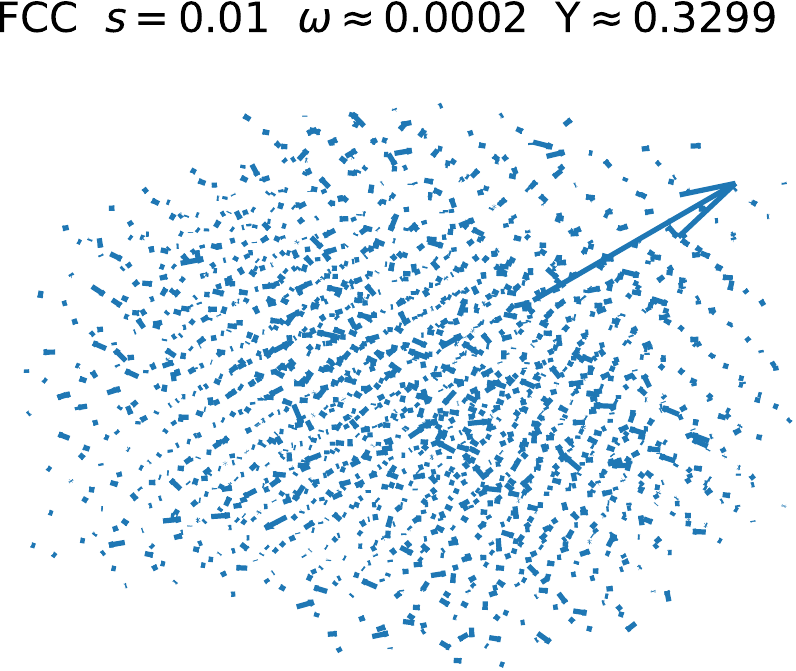}\\
\includegraphics[scale=.65]{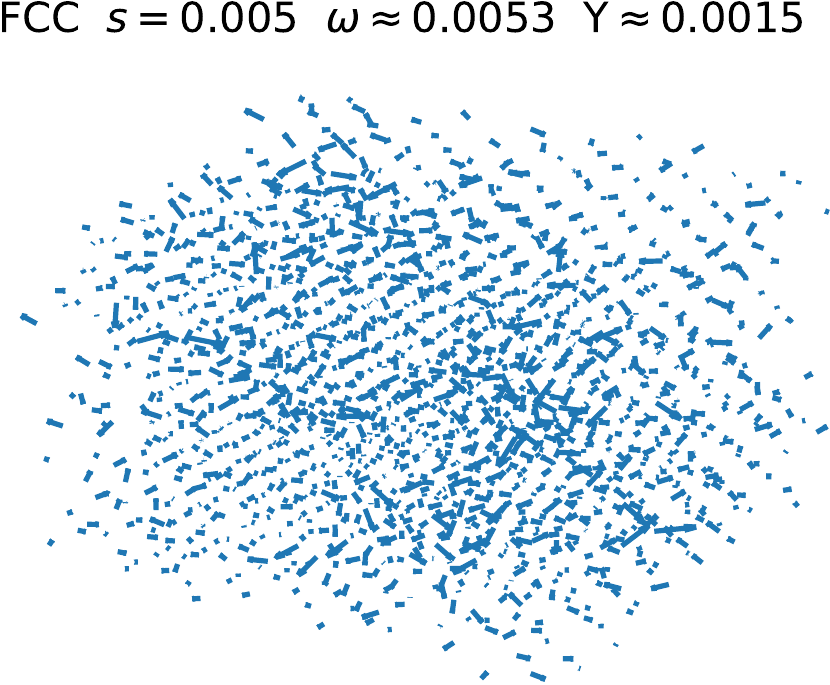}
\includegraphics[scale=.65]{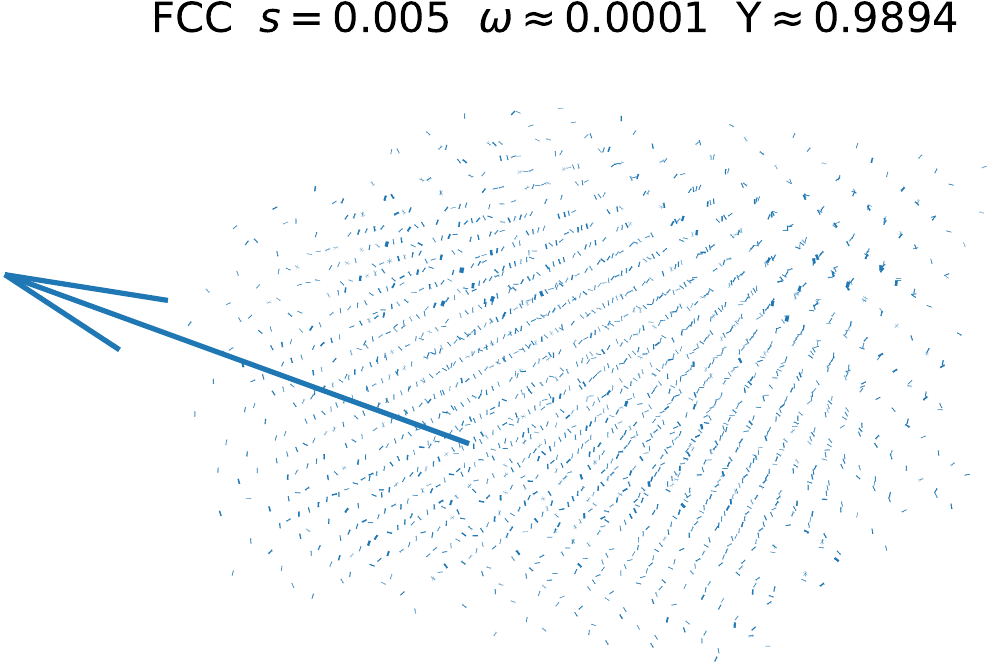}
\caption{ Plot of one of the smallest- (left column) and largest- (right column) IPR normal modes of a jammed 3d FCC near-crystal of $N = 2048$ particles for each polydispersity $s$. The frequency $\omega$ and the IPR $\mm{Y}$ are given (Fig.~\ref{fig:IPR}). 
}
\label{fig:nmodes3DFCC-A}
\end{figure*}

In order to gauge the amount of localization the Hessian eigenvectors exhibit we calculate the inverse participation ratio
\begin{eqnarray} 
\label{ipr}
Y(\omega) = \frac{\sum_i^N | \vec{u}_i(\omega)|^4}{[\sum_i^N | \vec{u}_i(\omega)|^2]^2},
\end{eqnarray}
This construction for the IPR has the property $Y \sim 1/N$ for extended eigenvectors and $Y \sim 1$ for localized eigenvectors~\cite{silbert_normal_2009, widmer-cooper_irreversible_2008, xu_anharmonic_2010, mizuno_elastic_2013, manning_vibrational_2011, charbonneau_universal_2016}.
In studies of amorphous jamming~\cite{charbonneau_universal_2016} it has been found that for small frequencies the IPR of maximally amorphous solids exhibits more localization at $\varphi$ higher than $\varphi_J$ while the smallest localization can be seen near the jamming point. Our results, Fig.~\ref{fig:IPR}, show that the more amorphised a near-crystal is it exhibits less localization for small frequencies at the jamming point in accordance with the behavior of amorphous packings~\cite{charbonneau_universal_2016}. However, for more ordered near-crystals at the onset of rigidity a greater localization is revealed in addition to the main, more delocalized, majority which itself displays a dimension-dependent scaling with frequency, Fig.~\ref{fig:IPR}. In SM~\cite{SI} Figs.~\ref{fig:nmodes2DHEX-A},\ref{fig:nmodes3DFCC-A} we plot the smallest- and largest-IPR normal mode for each polyispersity in $d=2,3$.

{\it Near-Crystal Crystalline Properties:}
We quantify crystalline order with measures of translational and bond-orientational order appropriately chosen for HEX and FCC~\cite{steinhardt_bond-orientational_1983, bocquet_amorphization_1992, tong_crystals_2015}. The translational order for both structures is measured by
\begin{eqnarray} 
\label{eq:Qt}
Q_{T} = \frac{1}{N} \left | \sum_{i=1}^N e^{ i\vec{G} \cdot \vec{r}_{i} } \right |  
\end{eqnarray}
where $\vec{G}$ is a reciprocal lattice vector and $\vec{r}_{i}$ is the location of particle $i$. The sum is over all $N$ particles including rattlers. We perform this sum for a total of $d$ distinct reciprocal lattice vectors and average the results. A common measure of the bond-orientational order for HEX in $d=2$ is 
\begin{eqnarray} 
\label{eq:Q6}
Q_{6} = \frac{1}{N_b} \left | {\sum_{i=1}^{N_b} e^{i6\theta_{i}}} \right |
\end{eqnarray}
where $\theta_{i}$ is the angle the $i$-th bond makes with the x-axis and $N_b$ is the total number of distinct pairs of neighbors in the packing even if they are not in contact including rattlers.

\begin{figure}[th]
\centering
\includegraphics[scale=.6]{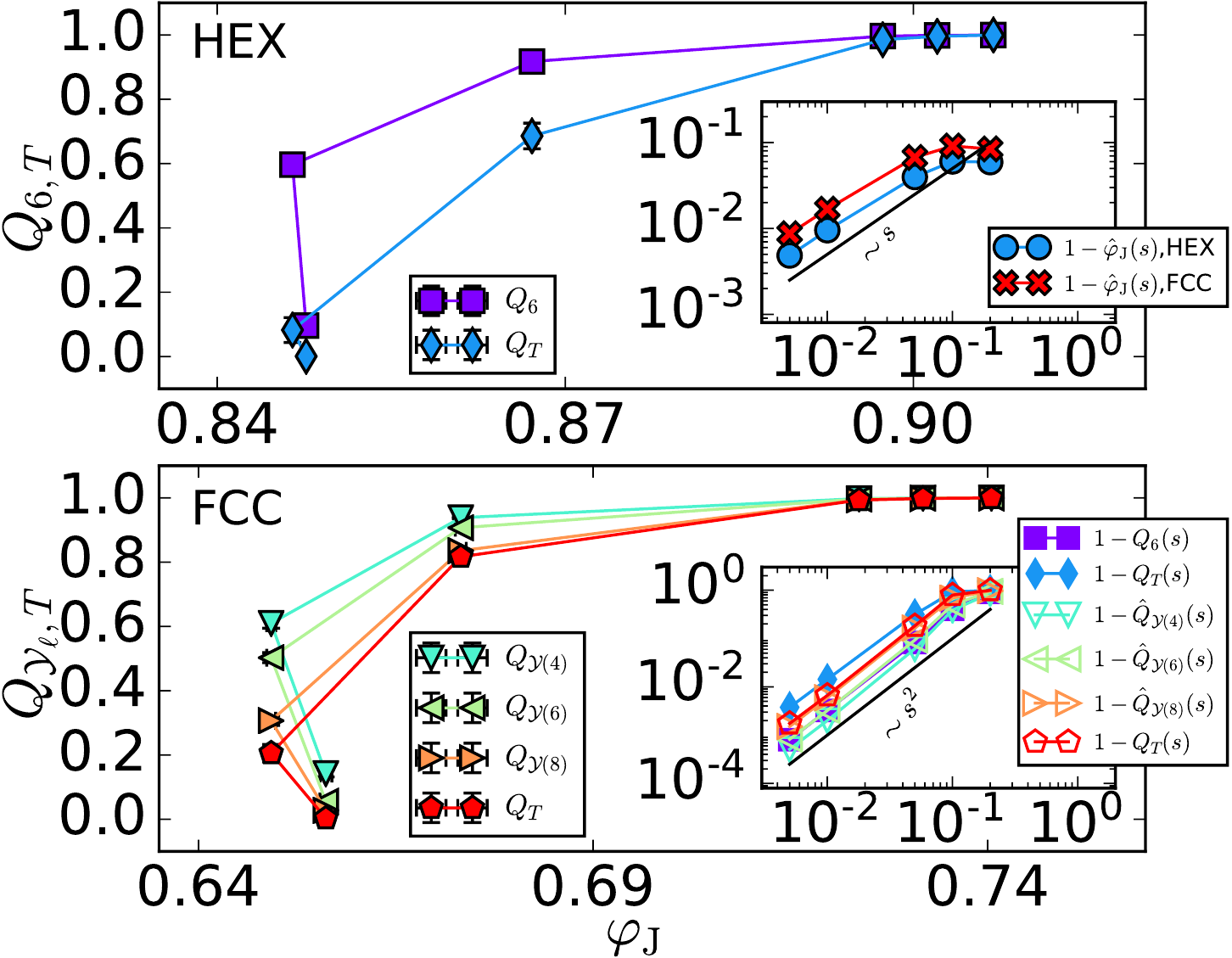}
\caption{Order-packing fraction plots at jamming for minimally polydisperse near-crystals. The points from high to low $\varphi_{\mm{J}}$, right-to-left, correspond to the polydispersities $s = 0.0, 0.005, 0.01, 0.05, 0.1, 0.2$. 
(top inset) Packing fraction at jamming and, (bottom inset) bond-orientational and translational order, for HEX and FCC near-crystals scale linearly, $\varphi_{\mm{J}}(s) \sim \varphi_{\mm{cp}}(1 - s)$, and parabolically, $Q_{x}(s) \sim Q_{x}(0)(1 - s^2)$, with polydispersity $s<s_{a}$. Symbols $\hat{Q}_{x}(s) = Q_{x}(s) / Q_{x}(0)$ for $x=\{T, 6, \mathcal{Y}(\ell)\}$ and  $\hat{\varphi}_{\mm{J}} = \varphi_{\mm{J}}(s) / \varphi_{\mm{cp}}$. }
\label{fig:OrdervsPhi}
\end{figure}

\begin{figure*}[]
\centering
\includegraphics[scale=.6]{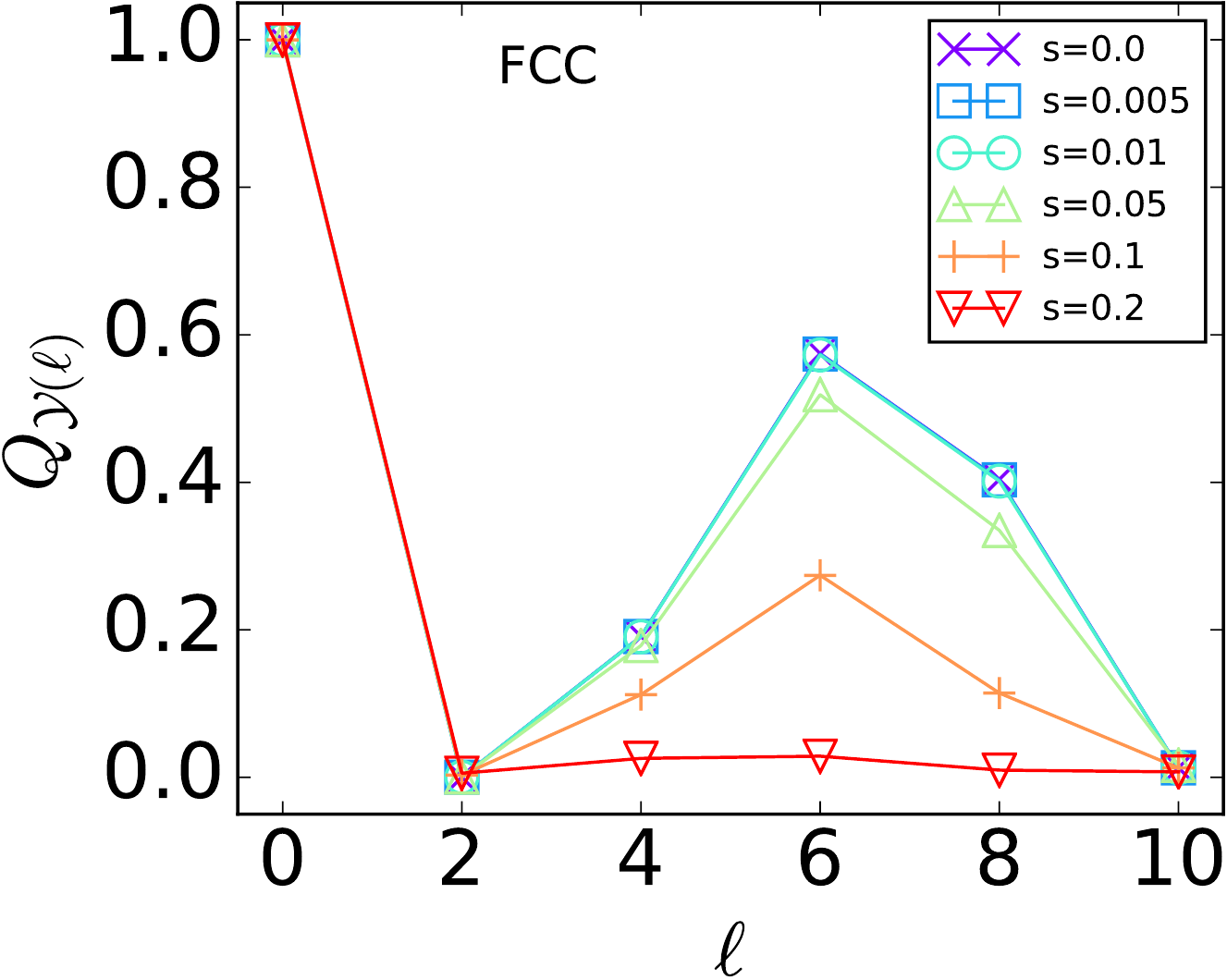}
\caption{Bond orientational order for FCC, isostatic for polydispersities $s = 0.005, 0.01, 0.05, 0.1, 0.2$, hyperstatic for $s=0.0$. The characteristic FCC profile weakens and eventually flattens out as polydispersity is increased towards maximum amorphisation (Fig.~\ref{fig:OrdervsPhi}).}
\label{fig:bondOrderFCC}
\end{figure*}

\begin{figure*}[]
\centering
\includegraphics[scale=.55]{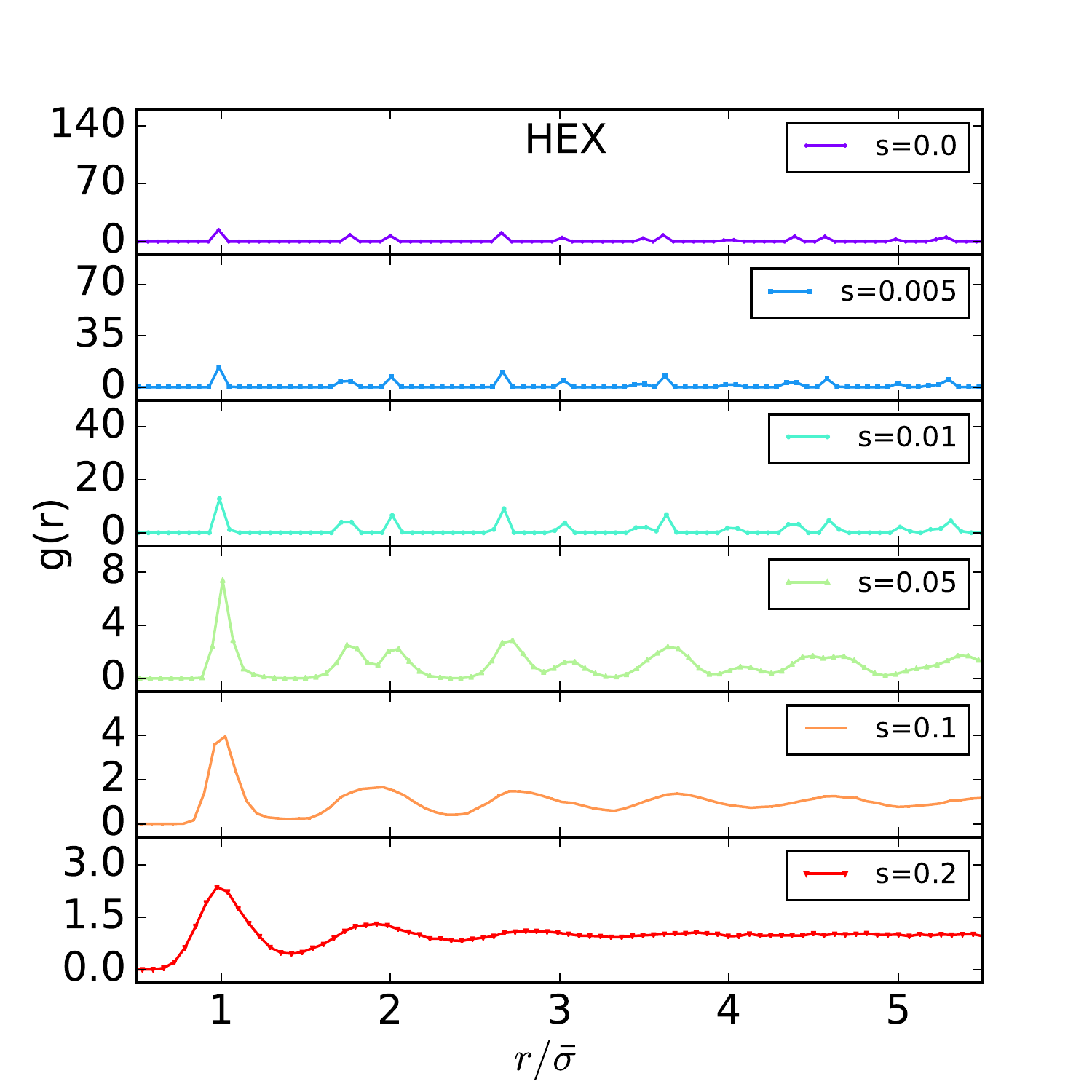}
\includegraphics[scale=.55]{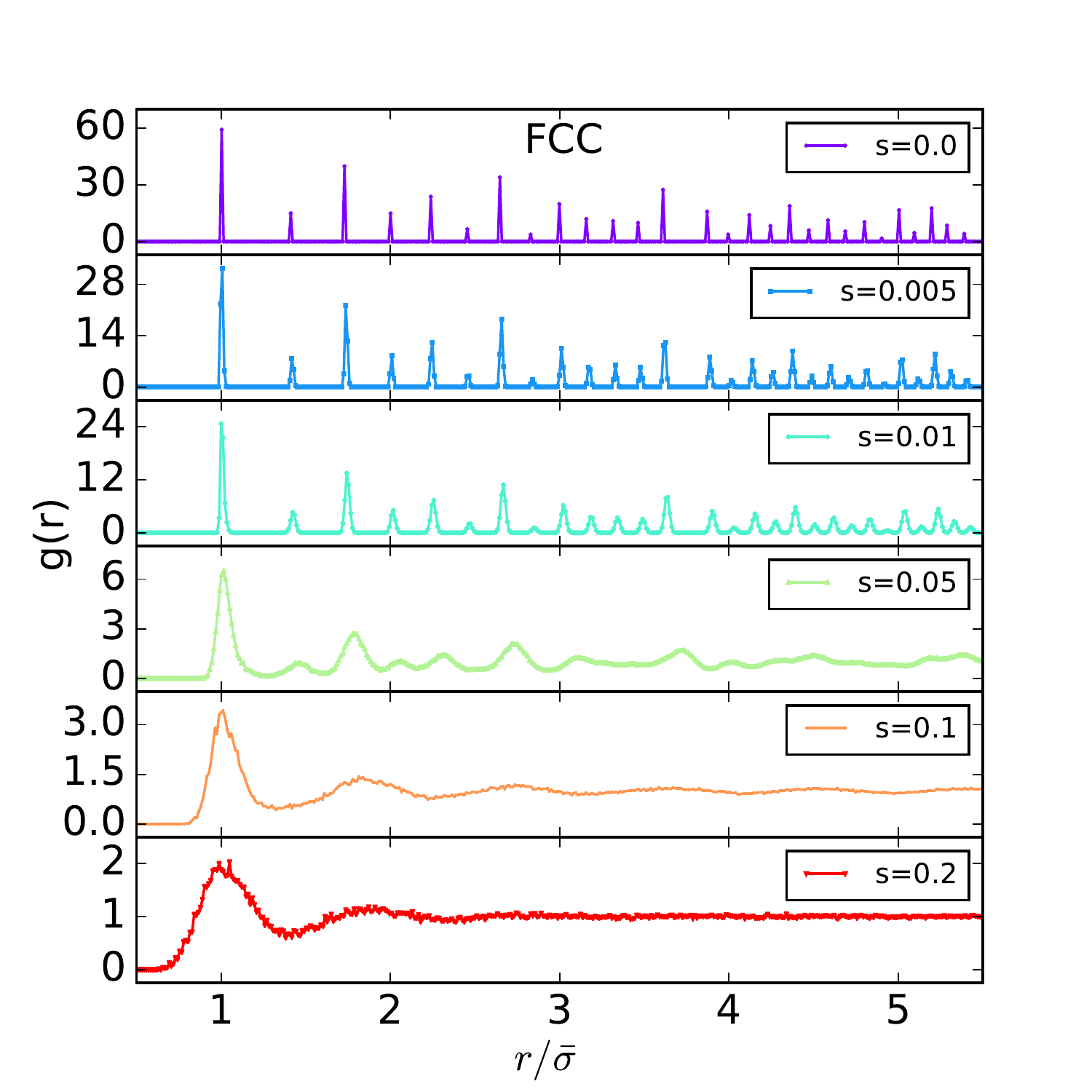}
\caption{ Pair correlation function for near-crystals, (left) HEX and (right) FCC, isostatic for polydispersities $s = 0.005, 0.01, 0.05, 0.1, 0.2$, hyperstatic for $s=0.0$. The sharp crystal peaks broaden and eventually the majority of them disappear, except for the peak at first neighboring contact, as polydispersity is increased towards maximum amorphisation (Fig.~\ref{fig:OrdervsPhi}). 
}
\label{fig:gofrHEXFCC}
\end{figure*}

In $d=3$ a bond direction can be characterized by its overlap with the spherical harmonic  $\mathcal{Y}_{\ell m}(\theta, \phi)$.
\begin{eqnarray} 
\label{eq:Qy}
Q_{\mathcal{Y}(\ell)} =  \left [ \frac{4 \pi}{2\ell +1}  \sum_{m=-\ell}^{\ell}  \left | \frac{1}{N_b} \sum_{i=1}^{N_b} \mathcal{Y}_{\ell m}(\theta_{i}, \phi_{i}) \right |^2 \right ]^{1/2}
\end{eqnarray}
shown for FCC in Fig.~\ref{fig:bondOrderFCC}~\cite{steinhardt_bond-orientational_1983}. In the absence of any polydispersity all measures of translational and bond-orientational order match the corresponding crystal values. As polydispesity is increased all measures of order decay parabolically with $s$~\cite{tong_crystals_2015}, 
\begin{eqnarray} 
Q_{x}(s) \sim Q_{x}(0)(1 - s^2)
\label{eq:Qx}
\end{eqnarray}
where $x=\{T, 6, \mathcal{Y}(\ell)\}$, see Figure \ref{fig:OrdervsPhi} (bottom inset). In Figure \ref{fig:OrdervsPhi} (top inset) we show that the packing fraction at jamming relates linearly with polydispersity 
\begin{eqnarray} 
\varphi_{\mm{J}}(s) \sim \varphi_{\mm{cp}}(1 - s)
\label{eq:PhiIso}
\end{eqnarray}
where $\varphi_{\mm{cp}}$ is the maximum packing fraction of the perfect crystal. Eqs.~\ref{eq:Qx} and ~\ref{eq:PhiIso} taken together quantify the amount of amorphisation and packing fraction of jammed near-crystals up to a suspected amorphisation point, $s_{a} \gtrsim 0.1$, echoing the amorphisation point $s'_{a} \sim 0.11$ found above $\varphi_{\mm{cp}}$~\cite{amorph1}. Beyond $s_{a}$ the packings are in a maximally amorphous phase~\cite{bocquet_amorphization_1992, mizuno_elastic_2013, tong_crystals_2015} as is depicted with global measures of order versus packing fraction in Fig.~\ref{fig:OrdervsPhi} and the spatially resolved pair distribution in Fig.~\ref{fig:gofrHEXFCC} in SM~\cite{SI}.

{\it Discussion and Conclusion:}
We have discovered that single-species, highly-symmetric, maximum-packing crystals, namely HEX in 2d and FCC in 3 dimensions, can reach an isostatic jamming point for a wide range of very small, numerically-achievable, amounts of polydispersity, $s \gtrsim 10^{-4}$. Some critical properties of isostatic near-crystals, such as $\theta_{\mm{ext}}$ and the flat, non-Debye scaling of VDOS, are the same as maximally amorphous packings for $d \ge 2$ and the infinite-dimensional MFT. Others, such as $\theta_{\mm{\ell oc}}$, $\gamma$, and the low-frequency IPR are dependent on polydispersity. Even though the jamming universality of near-crystals shares some prominent features with the jamming universality of maximally amorphous packings, the two may belong to distinct classes since they appear dissimilar in other defining aspects.

%These findings indicate that the jamming universality maximally amorphous packings may not be able to describe jammed near-crystals. 
%These findings indicate a non-equilibrium universality class for near-crystals that may be distinct from maximally amorphous jamming. 
%Of course this has to be checked with \lq\lq infinitely" more numeric data and analytically in the thermodynamic limit.

The packing fraction of the near-crystals at jamming decays linearly with $s$ below the maximum packing fraction of the perfect crystal in both dimensions $2$ and $3$, Eq.~\ref{eq:PhiIso}, signifying that the perfect crystal exists only at the complete absence of frustration~\cite{tong_crystals_2015, charbonneau_glassy_2019}. Combining Eq.~\ref{eq:PhiIso} with the decay of amorphisation, Eq.~\ref{eq:Qx}, we arrive at a parabolic relation of order versus jamming packing fraction. This finding can be compared with the work on the maximally random jammed state (MRJ)~\cite{torquato_jammed_2010, jiao_nonuniversality_2011, atkinson_existence_2014}, keeping in mind that different protocols were used and that the MRJ is defined for an individual configuration while the near-crystal results are ensemble-averaged and presumably involve a structural amorphisation transition~\cite{bocquet_amorphization_1992, mizuno_elastic_2013, tong_crystals_2015} at isotaticity. We are working on acquiring more data to capture and characterize point $s_{a}$ and will report on it in the near future.
%Near-crystalline jammed packings and the amorphisation transition are out-of-equilibrium phenomena that appear analogous with the equilibrium hexatic phase and the hexatic-to-liquid transition~\cite{kapfer_two-dimensional_2015, nelson_order_1982}. 
%\gt{cite 1 yodh exp, 1 nelson/halperin prb1979}
%\gt{remove the reentrant citation} ~\cite{nelson_reentrant_1983}
%\gt{???have to check for d=2 and find d=3 cubatic paper????}. forget cubatic as i have no papers on it
%\gt{maybe cite more KT HalperinNelson Young papers here OR just cite the KapferKrauth PRL2015 which has an intro review to the KTHNY scenario}

Experiments on molecular~\cite{chumakov_role_2014, szewczyk_glassy_2015, gebbia_glassy_2017, jezowski_glassy_2018} and colloidal~\cite{kaya_normal_2010} ordered systems have shown that crystals with minimal amounts or no apparent disorder can behave amorphously as depicted in the VDOS and the related heat capacity and thermal conductivity scaling at low temperature. Recently a theory with anharmonicity and dissipation was proposed to unify those findings with traditional glassy physics~\cite{baggioli_universal_2019}. Our work establishes that packings at the onset of rigidity and high order, not only possess a non-Debye VDOS but also exhibit extended-force statistics of the same universality as maximally amorphous packings and the related Gardner mean-field theory.
%that athermal jammed packings with long range order can reach isostaticity with minimal disorder and produce a flat, non-debye VDOS at the onset of rigidity and high order. It also establishes that the resulting force statistics for extended perturbations exhibit the same universal behavior as maximally amorphous packings and the related Gardner mean-field theory. 
On the other hand we show that the localized response of jammed near-crystals is more complicated than their maximally amorphous counterparts. Thus it will have to be investigated more thoroughly~(\cite{ono-dit-biot_rearrangement_2020}) in order to be incorporated in a complete theory~(\cite{ikeda2019jamming}).

%Our work contributes along this direction for the athermal packing problem of spheres with implications for condensed matter systems such as granular media, colloids, glasses, for biological structural phenomena such as cell patterning into tissue, macro-molecular crowding in cells, animal herding or organizing into colonies, for information theory such as constraint satisfaction, optimization and machine learning problems and, digitization, propagation and error-correction of signals.
%problems (liquids/crystals, colloids, granular media, powders) but also for information theory problems (signal digitization, error-correction, optimization).

\emph{Acknowledgements:}
%We would like to thank S. Katira and J. Toner for inspiration. We would like to thank F.S., P.C., E.C., G.P. H.Ikeda, for useful discussions.
I would like to thank S. Katira, J. Toner, F. Spaepen, P. Charbonneau, E. Corwin, J. Dale, M. van der Naald, G. Sicuro, G. Parisi, H. Ikeda for useful discussions. I would like to thank P. Baldan, R. Diaz, G. Folena, S. Franchini for a critical reading of the manuscript and suggestions on presentation and citations. This work has been supported in part by the Simons Foundation (grant No. 454949, G. Parisi and, grant No. 454939, E. Corwin).

\bibliography{five}

\newpage
\clearpage

\beginsupplement

\section{Supplementary Material}

We define packing fraction commonly as 
\begin{eqnarray}
\varphi= \frac{\Sigma_{i=1}^{N} v_i}{\Pi_{\alpha=1}^{d}L_{\alpha}}
\end{eqnarray}
$v_{i}=V_{d}(\sigma_{i}/2)^{d}$ is the volume in $d$ dimensions of particle $i$, $\sigma_{i}$ its diameter,  and $L_{\alpha}$ the $\alpha$-th side of the simulation box. $V_{d}=\frac{\pi^{\frac{d}{2}}}{\frac{d}{2}\Gamma(\frac{d}{2})}$ and $\Gamma(x)$ the Gamma function. 

The perfect HEX crystal in $d=2$ packs maximally ($s=0$) at 
\begin{eqnarray}
\varphi_{\mm{cp},\mm{HEX}}=\frac{\sqrt{3} \pi}{6} \approx 0.90690
\end{eqnarray}
while the FCC in $d=3$ at zero polydispersity has 
\begin{eqnarray}
\varphi_{\mm{cp},\mm{FCC}}=\frac{\pi}{3\sqrt{2}} \approx 0.74048
\end{eqnarray} 
The simulation box is commensurate with the perfect crystal packing, i.e. for FCC $L_{x}=L_{y}=L_{z}$ while for HEX $L_{x}=\frac{\sqrt{3}}{2}L_{y}$.

Polydispersity is introduced through the lognormal distribution. Starting from equal radii in the initial perfect crystal we multiply each radius with a random number drawn from the log-normal distribution with unit mean and variance $s^2$. 

%Unifrom random numbers we get from the Mersene-Twister algorithm.

\begin{figure*}[]
\centering
\includegraphics[scale=.86]{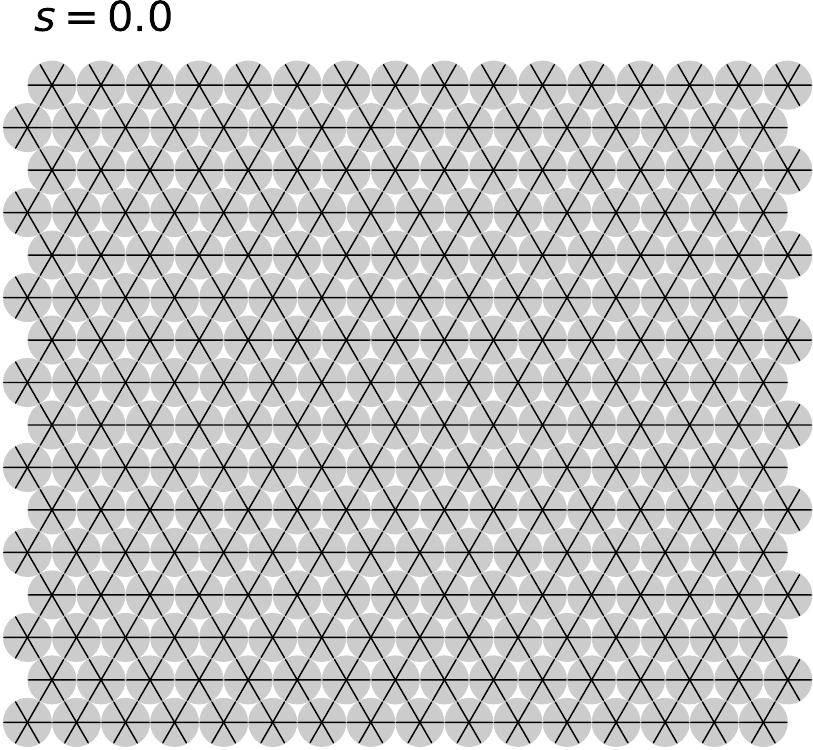}
\includegraphics[scale=.86]{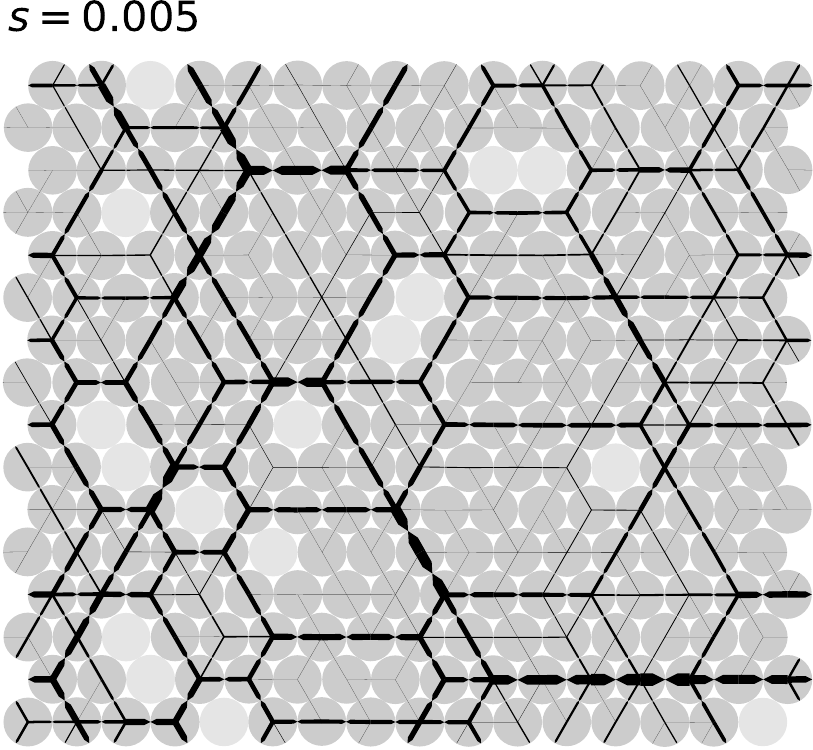}
\includegraphics[scale=.86]{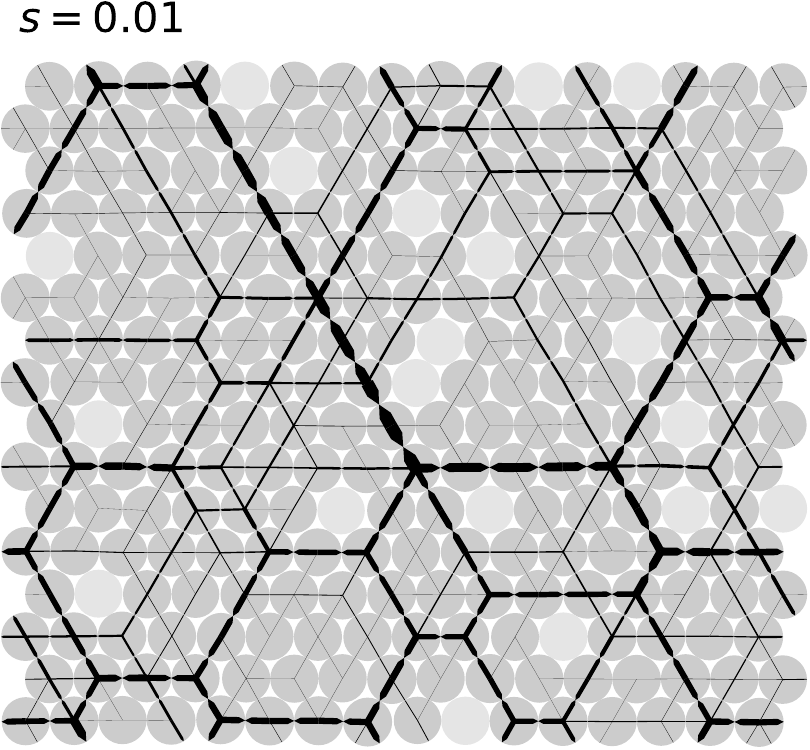}
\includegraphics[scale=.86]{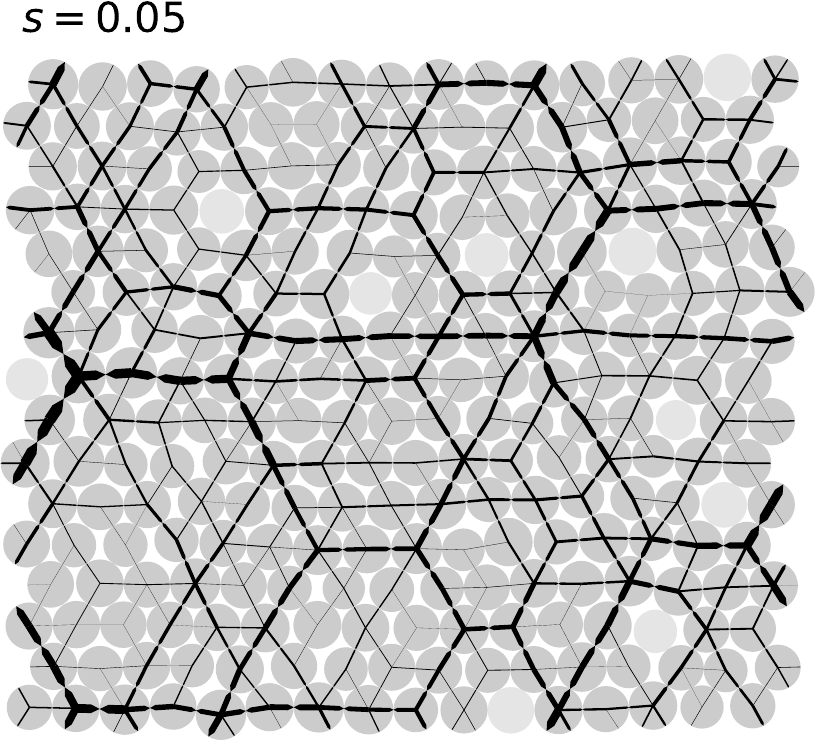}
\includegraphics[scale=.90]{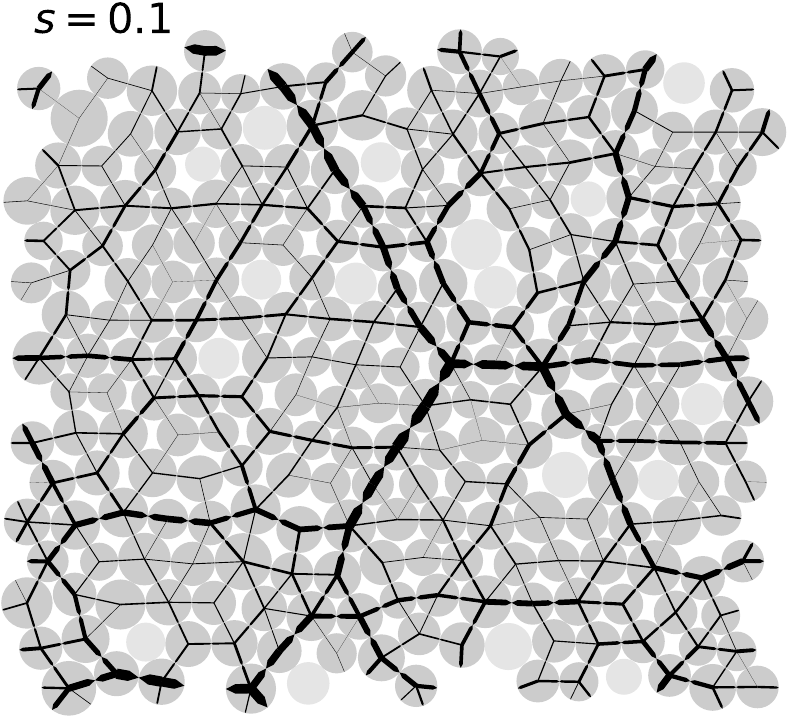}
\includegraphics[scale=.89]{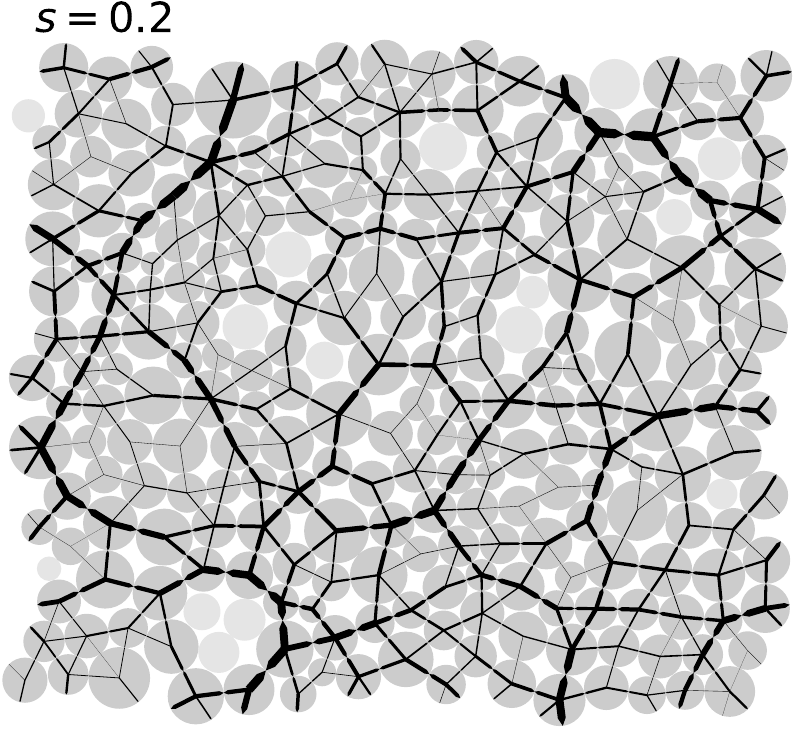}
\caption{ Visualization of a 2d HEX near-crystal of $N = 256$ particles at polydispersity $s$, isostatic for $s>0.0$, hyperstatic for $s=0.0$. For very small polydispersity a near-crystalline packing can reach a jamming point as can be evidenced in the disordered force network.\\
(top left) $s = 0.0$, $\varphi \approx 0.90690 \approx \varphi_{\mm{cp}}$. The $N_{c} = 768$ forces are sketched to be all equal. There are no rattlers ($N_{\mm{r}}=0$).\\
(top right) $s = 0.005$, $\varphi \approx 0.90260 \approx \varphi_{\mm{iso}} \approx 0.995 \varphi_{\mm{cp}}$,  $N_{c} = 479 \equiv N_{\mm{c}, \mm{iso}}$ forces and, $N_{\mm{r}}=16$ rattlers (\lq\lq lighter" particles).\\
(mid left) (Identical to Fig.~\ref{fig:visualize2DHEX2}) $s = 0.01$, $\varphi \approx 0.89801 \approx \varphi_{\mm{iso}} \approx 0.990 \varphi_{\mm{cp}}$, $N_{c} = 473 \equiv N_{\mm{c}, \mm{iso}}$ forces and, $N_{\mm{r}}=19$ rattlers.\\
(mid right) $s = 0.05$, $\varphi \approx 0.86679 \approx \varphi_{\mm{iso}} \approx 0.956 \varphi_{\mm{cp}}$,  $N_{c} = 491 \equiv N_{\mm{c}, \mm{iso}}$ forces and, $N_{\mm{r}}=10$ rattlers.\\
(bottom left) $s = 0.1$, $\varphi \approx 0.84626 \approx \varphi_{\mm{iso}} \approx 0.933 \varphi_{\mm{cp}}$, $N_{c} = 477 \equiv N_{\mm{c}, \mm{iso}}$ forces and, $N_{\mm{r}}=17$ rattlers.\\
(bottom right) $s = 0.2$, $\varphi \approx 0.84556 \approx \varphi_{\mm{iso}} \approx 0.932 \varphi_{\mm{cp}}$, $N_{c} = 481 \equiv N_{\mm{c}, \mm{iso}}$ forces and, $N_{\mm{r}}=15$ rattlers.  
}
\label{fig:vis2DHEX012345}
\end{figure*}

\begin{figure*}
\includegraphics[width=0.76\columnwidth]{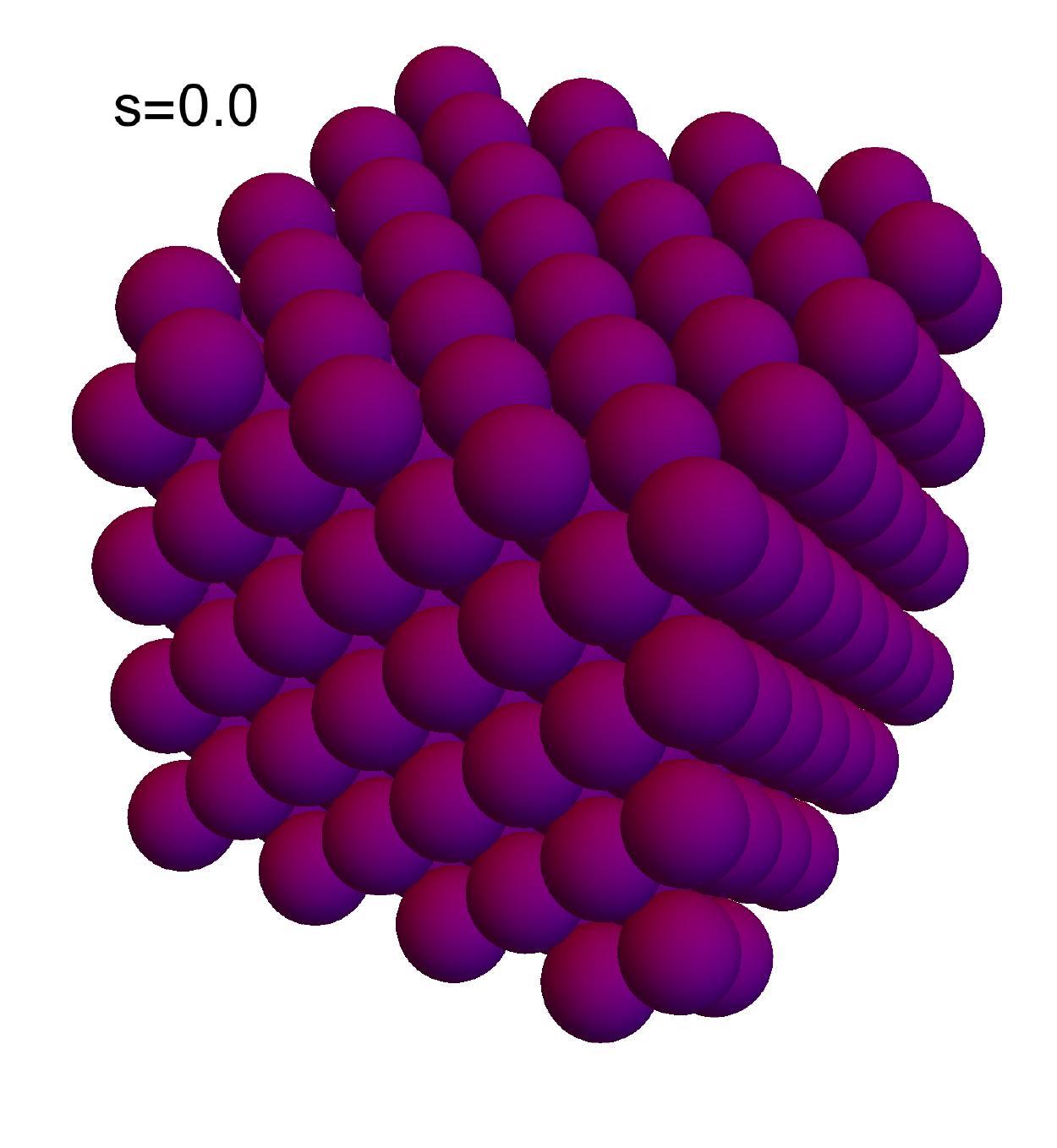}\\
\includegraphics[width=0.90\columnwidth]{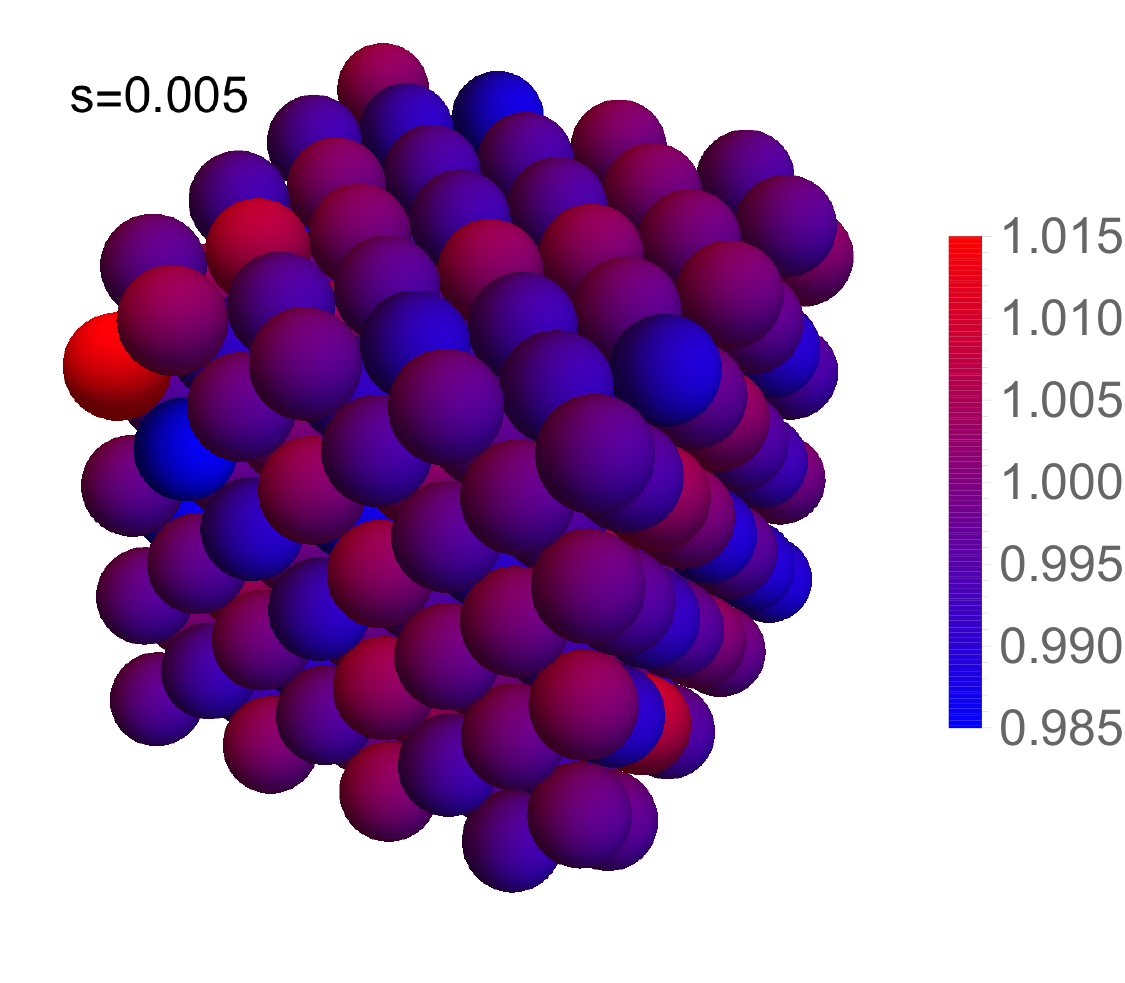}
\includegraphics[width=0.90\columnwidth]{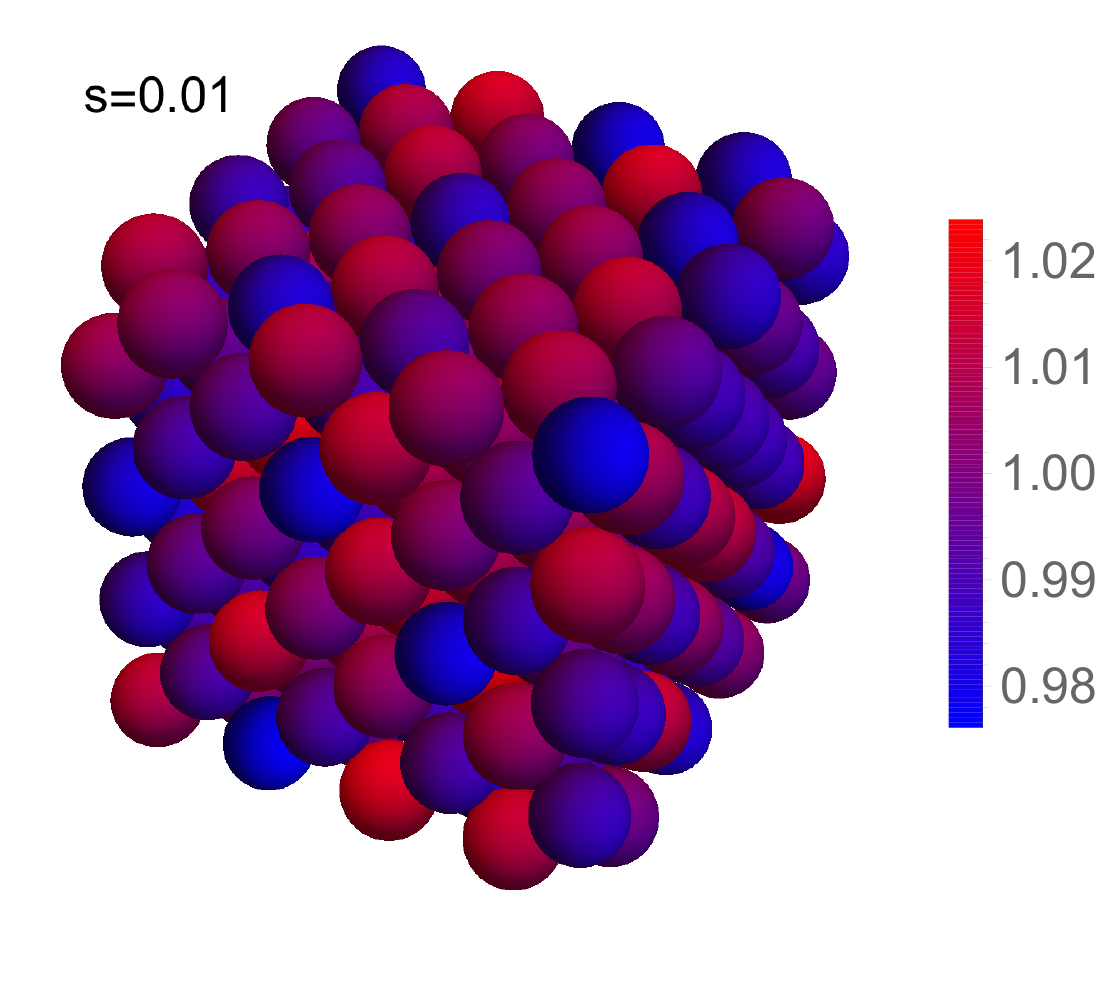}\\
\includegraphics[width=0.90\columnwidth]{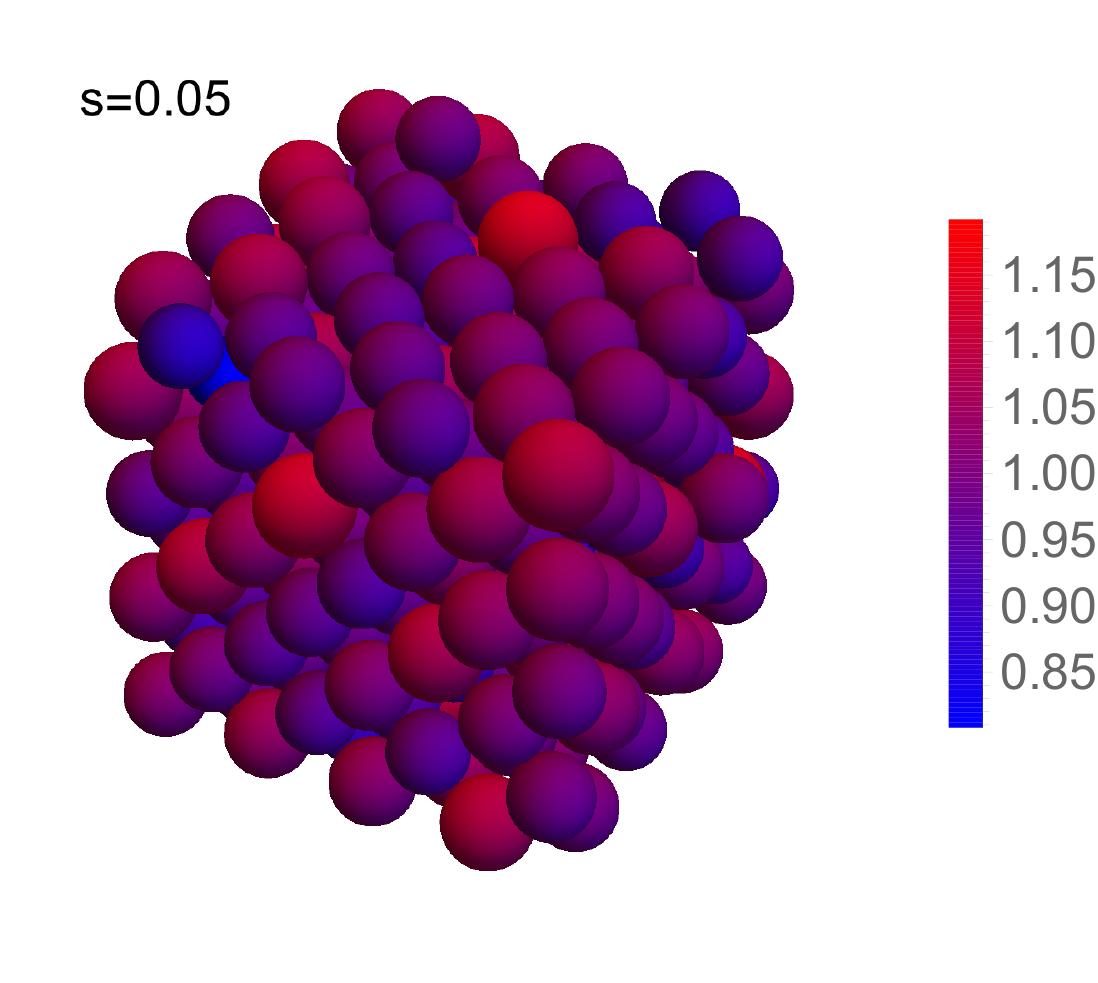}
\includegraphics[width=0.90\columnwidth]{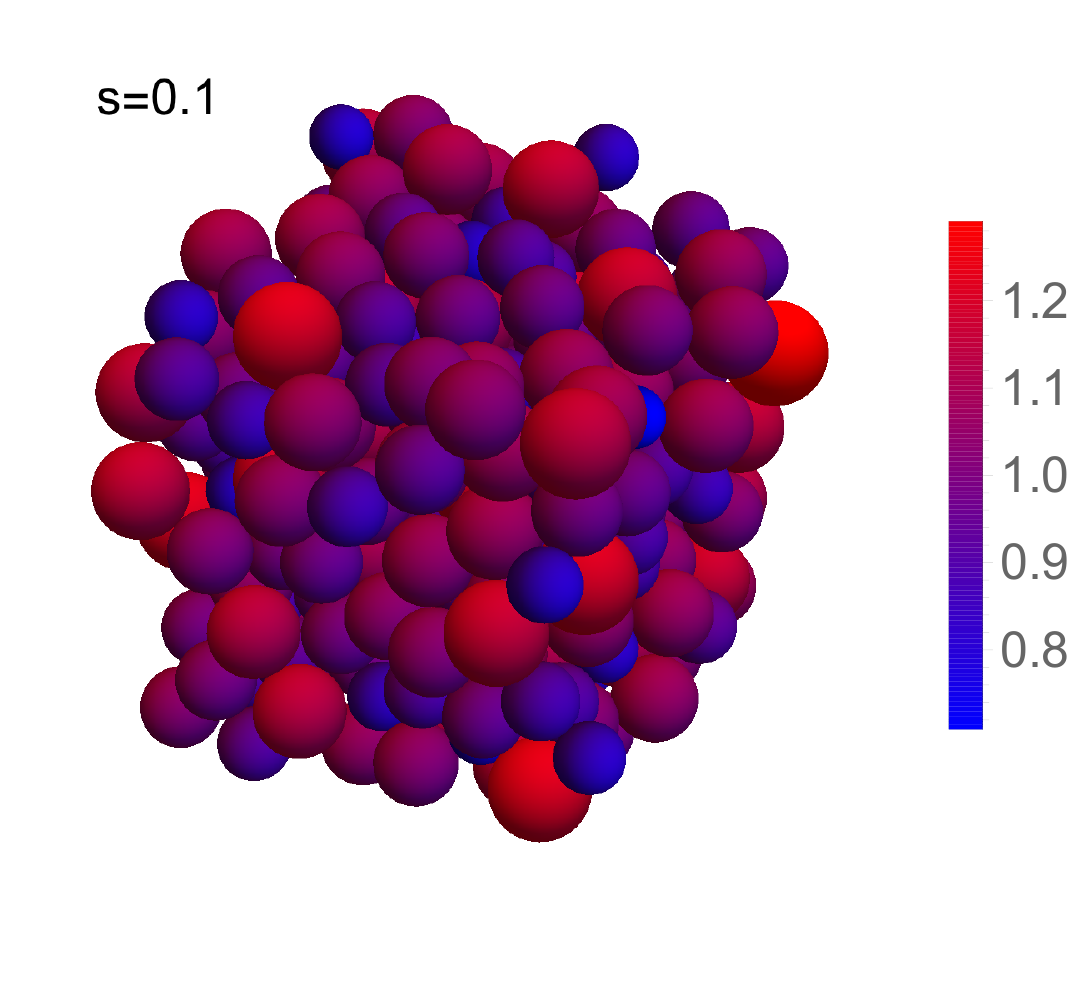}
\caption{ Visualization of a 3d FCC near-crystal of $N = 256$ particles at various polydispersities $s$, isostatic for $s>0.0$, hyperstatic for $s=0.0$. The colored scale bar indicates the relative patricle poyldispersity. (Rattles are shown but not indicated separately.) \\
i) $s=0.0$, $\varphi \approx 0.74048 \approx \varphi_{\mm{cp}}$. $N_{c} = 1536$ contacts and,  $N_{\mm{r}} = 0$ rattlers.\\
ii) $s = 0.005$, $\varphi \approx 0.73116 \approx \varphi_{\mm{iso}} \approx 0.987 \varphi_{\mm{cp}}$. $N_{c} = 703 \equiv N_{\mm{c}, \mm{iso}}$ contacts and,  $N_{\mm{r}} = 21$ rattlers. \\
iii) $s = 0.01$, $\varphi \approx 0.72157 \approx \varphi_{\mm{iso}} \approx 0.974 \varphi_{\mm{cp}}$. $N_{c} = 703 \equiv N_{\mm{c}, \mm{iso}}$ contacts and,  $N_{\mm{r}} = 21$ rattlers. \\
iv) $s = 0.05$, $\varphi \approx 0.67176 \approx \varphi_{\mm{iso}} \approx 0.907 \varphi_{\mm{cp}}$. $N_{c} = 724 \equiv N_{\mm{c}, \mm{iso}}$ contacts and,  $N_{\mm{r}} = 14$ rattlers. \\
v) $s = 0.1$, $\varphi \approx 0.64837 \approx \varphi_{\mm{iso}} \approx 0.876 \varphi_{\mm{cp}}$. $N_{c} = 739 \equiv N_{\mm{c}, \mm{iso}}$ contacts and,  $N_{\mm{r}} = 9$ rattlers.\\
}
\label{fig:vis3DFCC012}
\end{figure*}

\end{document}